\documentclass[twocolumn,prl,superscriptaddress,floatfix,showpacs]{revtex4-1}
\usepackage[utf8]{inputenc}
\usepackage{amsmath}
\usepackage{amssymb}
\usepackage{graphicx}
\usepackage{accents}

\makeatletter
\usepackage{graphics}
\usepackage{epsfig}
\usepackage{color}

\newcommand{\be}{\begin{equation}} \newcommand{\ee}{\end{equation}}
\newcommand{\bea}{\begin{eqnarray}} \newcommand{\eea}{\end{eqnarray}}

\makeatother

\begin{document}

\title{A robust balancing mechanism for spiking neural networks}

\author{Antonio Politi}
\affiliation{Institute for Complex Systems and Mathematical Biology and Department of Physics, Aberdeen AB24 3UE, UK}
\affiliation{CNR - Consiglio Nazionale delle Ricerche - Istituto dei Sistemi Complessi, via Madonna del Piano 10, 50019 Sesto Fiorentino, Italy}

\author{Alessandro Torcini}
\email[]{Corresponding Author: alessandro.torcini@cyu.fr}
\affiliation{Laboratoire de Physique Th\'eorique et Mod\'elisation, CY Cergy Paris Universit\'e, CNRS, UMR 8089,
95302 Cergy-Pontoise cedex, France}
\affiliation{CNR - Consiglio Nazionale delle Ricerche - Istituto dei Sistemi Complessi, via Madonna del Piano 10, 50019 Sesto Fiorentino, Italy}
\affiliation{INFN Sezione di Firenze, Via Sansone 1, 50019 Sesto Fiorentino, Italy}

\date{\today}

\begin{abstract}
Dynamical balance of excitation and inhibition is usually invoked to explain the irregular low firing activity observed in the cortex. We propose a robust 
nonlinear balancing mechanism for a random network of spiking neurons, which works also in absence of strong external currents. 
Biologically, the mechanism exploits the plasticity of
excitatory-excitatory synapses induced  by short-term depression. 
Mathematically, the nonlinear response of the synaptic activity 
is the key ingredient responsible for the emergence of a stable balanced regime. Our claim is supported by a simple self-consistent analysis accompanied by extensive simulations 
performed for increasing network sizes.
The observed regime is essentially fluctuation driven and characterized by highly irregular
spiking dynamics of all neurons.
\end{abstract}

\maketitle

{\bf  Neurons in the cortex fire irregularly and with a
low firing rate despite being subject to a continuous
stimulation (bombardment) from thousands of pre-synaptic
neurons. This seemingly counter-intuitive evolution 
has been explained in terms of the theory of dynamical
balance of excitation and inhibition \cite{bal1}, considered as 
one of the major contributions of theoretical physics to neuroscience.
However, this theory has been recently criticized, because it requires unphysically
large external currents, experimentally unjustified. 
Here, we propose a nonlinear balancing mechanism based on a biologically plausible form of synaptic plasticity 
(short-term depression) which works also  with weak external currents.}

Inferring the collective behavior of large ensembles of oscillators is a highly
challenging task, since it requires combining concepts and tools of nonlinear
dynamics with those of statistical mechanics
~\cite{kuramoto2012}.
Already the classification of proper ``thermodynamic" phases and of the conditions
for their emergence is a non trivial task: what are the qualitative differences
among the various regimes?
A preliminary difficulty is posed by the identification of appropriate model 
classes. Typically (but not exclusively), the coupling is assumed to result from
the linear composition of two-body interactions. However, already within this
simplified setup, a question arises in the case of massive coupling, when
the number $K$ of connections is proportional to the number $N$ of oscillators
(mean-field models being the ultimate example).
In fact, a meaningful thermodynamic limit requires the coupling term to be finite 
for $K\to\infty$. This is typically ensured by assuming a coupling
constant of order $1/K$: we call this type-I coupling and
the Kuramoto model is perhaps the most famous example \cite{acebron2005}.
An alternative approach can be adopted, when the single two-body coupling terms
are, on average, equal to zero. In such cases, it makes sense to assume that
the coupling constant is of order $1/\sqrt{K}$. We call it type-II coupling and
the XY spin-glass model \cite{sherrington1975,kirkpatrick1978}
is one of the most famous representatives of this class.

The Hamiltonian-mean-field \cite{antoni1995} is an enlightening example, which encompasses both options. This extension of the Kuramoto model, if equipped with type-I coupling, proves useful to describe chaotic properties of the synchronized (magnetized) phase \cite{ginelli2011}; if equipped with type-II coupling, it has helped to discover nontrivial properties of the asynchronous (unmagnetized) regime \cite{konishi1990}. 

A paradigmatic system where oscillator networks find application is represented by
the mammalian brain, whose dynamics follows from a complex interplay between
microscopic (single neuron) and macroscopic features.
In particular, pulse-coupled phase oscillators ~\cite{winfree,strogatz1993,pikovsky2015}
, purposely introduced to describe neuronal dynamics, reproduce a large variety of phenomena~\cite{abbott1993, van1996, timme2002, vogels2005, haken2006,olmi2011}.

In this context, for mean-field models of globally coupled oscillators with type-I coupling, the stationary regime is often found to be asynchronous,  i.e. characterized by constant collective features (such as the local field potential) possibly accompanied by tiny fluctuations resulting from the finiteness of the neuronal population.
Partial synchrony may manifest itself as either periodic macroscopic oscillations~\cite{van1996,wang1996}, 
or irregular fluctuations~\cite{luccioli2010,ullner2016}.
Anyway, the corresponding single-neuron firing activity is typically regular 
(the coefficient of variation (CV) of the interspike intervals is small), 
contrasting the experimental evidence that cortical neurons {\it in vivo} operate erratically and with a relatively low
firing rate~\cite{softky1993} in spite of receiving stimulations from thousands of 
pre-synaptic neurons~\cite{destexhe1999,bruno2006,lefort2009}.   

An irregular firing activity is generated if the neurons operate in the so-called
fluctuation-driven regime, when they stay in proximity of the firing threshold, crossed at random  times thanks to self-generated fluctuations~\cite{shadlen1994}.
This can happen when inhibition is strong and accompanied by a random connectivity
which suppresses coherence across the neuronal population \cite{brunel1999}.

Altogether, it is widely accepted that all these features can dynamically emerge
whenever the underlying dynamics is in the so-called
{\it balanced} regime~\cite{bal1}, observed in excitatory-inhibitory networks characterized
by type-II coupling, an assumption consistent with optogenetic experiments {\it in vitro} \citep{barral2016}. A balanced state can be, e.g., found, by assuming:
(i) a sufficiently large in-degree $K$; 
(ii) coupling strengths of order $1/\sqrt{K}$; 
(iii) external currents of order $\sqrt{K}$~\cite{bal1,bal2,wolf,bal3,bal4,rosenbaum2014,pyle2016,matteo,noi}. 
If the external currents are of order $\mathcal{O}(1)$, excitation and inhibition still balance each other, but
the firing activity decreases as $1/\sqrt{K}$, suggesting that 
strong external currents  are a necessary ingredient. 
However, this latter hypothesis has recently received several criticisms~\cite{ahmadian2021,khajeh2022}
based on experimental evidences that the external inputs are ${\mathcal O}(1)$~\cite{chung1998,finn2007,lien2013}. 

In this paper, we show that the introduction of nonlinearities
can robustly sustain and stabilize a balanced regime,
where the irregular firing of excitatory and inhibitory neurons compensate each other,
without neither the inclusion of strong external currents, nor the {\it ad hoc} adjustment of parameter values.
The nonlinear mechanism, herein invoked is the well known
short-term synaptic depression (STD)~\cite{tsodyks2013},
arising from the finitude of available resources~\cite{tsodyks1997}.
It has been shown that depression has a prevalent effect on excitatory synapses in the visual cortex,
inducing dynamical variations of the balance between excitation and inhibition~\cite{varela1999}.

More precisely, we consider a plastic network of pulse-coupled phase-oscillators, where STD modifies
nonlinearly the synaptic inputs.
For the sake of simplicity and consistently with experimental indications \cite{varela1997,varela1999}, 
STD is assumed to act only on the synapses connecting excitatory neurons.
We show that this little adjustment suffices to ensure the self-sustainement of an irregular activity.

\paragraph{The model} 

We consider two coupled populations each composed of $N$ neurons. 
The evolution of the membrane potential $v_j^{e / i}$ of the $j$-th neuron within the
excitatory/inhibitory population follows from the equation,
\begin{equation}
	\dot v^{e/i}_j = F (v^{e/ i}_j) + G^\prime H(v^{e/ i}_j) C_j^{e/ i} \enskip, \quad v_j^{e/i} \in (-\infty;1] \enskip .
\label{eq:model00} 
\end{equation}
Whenever $v^{e/i}_j$ reaches the threshold 1, it is reset to 0, and, simultaneously, a smooth
post-synaptic $\alpha$-pulse $p_\alpha (t) = \alpha^2 t {\rm e}^{-\alpha t}$ is delivered to all the connected neurons, mimicking
a non-istantaneous synaptic transmission \citep{abbott1993,van1994,coombes2003}.
For large $\alpha$-values, $p_\alpha$ is well approximated by a $\delta$-pulse \citep{van1996}.
$F(v) > 0$ describes the bare neuron velocity field 
under the action of a weak constant external current, such that
it operates slightly supra-threshold; $C^{e/i}_j$ denotes the incoming synaptic recurrent current 
(see below for its definition); 
$H(v)$ gauges the impact of the current, which may depend on the value of the membrane potential;
finally, $G^\prime$ denotes the overall coupling strength.
For the sake of simplicity, $F(v)$ and $H(v)$ are taken to be the same for all neurons; the difference between excitatory
and inhibitory neurons is encoded in their mutual couplings.

Without loss of generality, $F(v)$ can be assumed to be constant. In fact, if we introduce the new variable $\phi$
obtained by integrating the ode $d \phi/d v = \omega/F(v)$ (under the condition $\phi(0) = 0$), Eq.~(\ref{eq:model00}) rewrites as
\begin{equation}
	\dot \phi^{e/i}_j = \omega + GZ(\phi^{e/ i}_j) C_j^{e/ i} \; .
\label{eq:model0} 
\end{equation}
where $Z(\phi) = \eta \omega H(v(\phi))/F(v(\phi))$ and $\eta$ is a normalization constant suitably chosen
to set the maximum of $Z(\phi)$ equal to  one (hence, $G = G^\prime/\eta$).
The value of $\omega$ is determined by imposing $\phi(1) = 1$
(this condition is equivalent to $\omega = 1/T$, where $T$ is the period of the bare neuron activity).
Hence, $\phi$ is a phase-like variable, while $Z(\phi)$ can be viewed as an effective phase response curve (PRC)
\cite{mirollo1990,timme}.

\begin{figure}
\begin{centering}
\includegraphics[width=0.4\textwidth,clip=true]{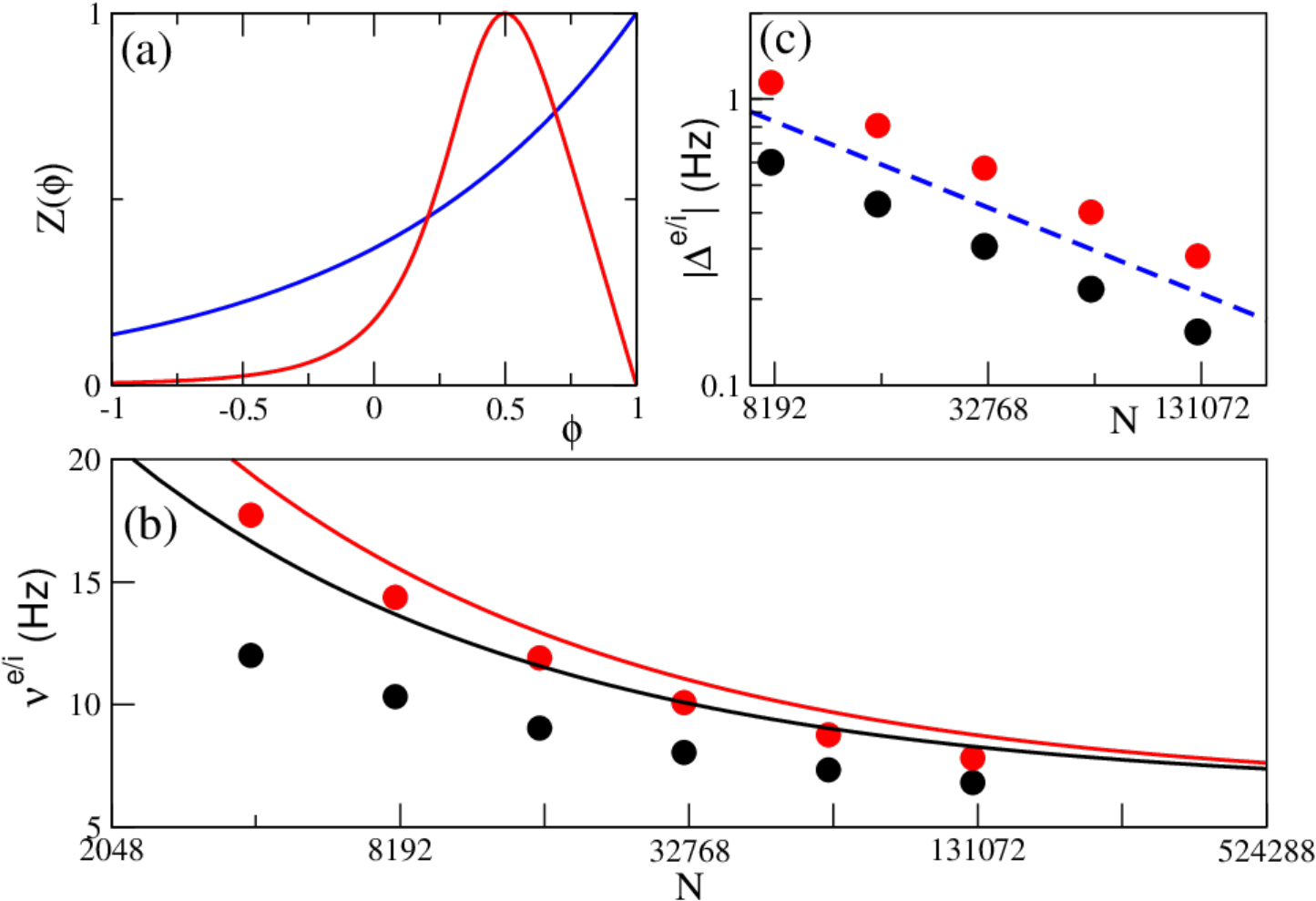}
\caption{(a) The PRC $Z(\phi)$ vs the phase-like variable $\phi$ for $Z_{\rm I}$ (red curve)
and $Z_{\rm LIF}$ (blue curve); (b) the population firing rates $\nu^e = \langle E^i_j\rangle$ and 
$\nu^i = \langle I_j \rangle$ versus $N$;
(c) the average unbalance $|\Delta^{e/i}|$  versus $N$, the blue dashed line denotes a power law decay as $1/\sqrt{N}$.
In (b-c) the black (red) color refers to excitatory (inhibitory) neurons and
the solid lines in (b) to the self-consistent approximations. The results
in (b-c) refer to $Z_{\rm I}$.
}
\label{fig:fig1}
\end{centering}
\end{figure}

Leaky Integrate-and-Fire (LIF) neurons are among the most popular models used in computational neuroscience 
(see, e.g. \cite{burkitt2006}). For current based coupling, they are characterized by $F_{\rm LIF}(v) = a - v$ 
(with $a>1$) and $H_{\rm LIF}(v) = 1$. 
If we introduce $\phi = \omega \ln [a/(a-v)]$ where $\omega = [\ln[a/(a-1)]^{-1}$, the LIF
model can be recast in the equivalent representation (\ref{eq:model0}), where
$Z_{\rm LIF}(\phi) = \mathrm{e}^{(\phi-1)}$ (see the blue curve in Fig.~\ref{fig:fig1}(a)).

In this paper, we have considered also
$Z_{\rm I} (\phi) = 12(1-\phi)/[5+(2-2\phi)^6]$ (see the red curve in Fig.~\ref{fig:fig1}(a)) for its continuity
at threshold, as usually assumed in realistic PRCs \cite{smeal2010}, 
and its resemblence to PRC for type I membrane excitability \cite{ermentrout1996}. 

Finally, the incoming synaptic currents are defined as
\begin{eqnarray}
C^{e}_j &\equiv& \frac{g^{e}_e}{\sqrt{K^e}} \mathcal{E}^{e}_j - \frac{g^{e}_i}{\sqrt{K^i}} \mathcal{I}_j \quad ,
\nonumber \\
C^{i}_j &\equiv& \frac{g^{i}_e}{\sqrt{K^e}} \mathcal{E}^{i}_j - \frac{g^{i}_i}{\sqrt{K^i}} \mathcal{I}_j 
\label{eq:C}
\end{eqnarray}
where the coefficients $(g^{e}_{e},g^e_i,g^i_e,g^i_i)$ quantify the specific intra and inter synaptic strengths 
of excitatory and inhibitory populations, while $K^{e/i}$ is the average in-degree, 
and $\mathcal{I}_j$ and $\mathcal{E}^{e/i}_j$ represent the incoming inhibitory and (effective) excitatory fields.
The inhibitory field obeys the differential equation
\begin{equation}
	\ddot{ \mathcal{I}}_j + 2\alpha \dot{\mathcal{I}}_j + \alpha^2 \mathcal{I}_j 
= \alpha^2  \!\!\!\! \sum_{n, m|t^i(n,m)<t}^\prime \delta (t-t^i(n,m)) \, ,
\label{field_i}
\end{equation}
where $\alpha$ is the inverse pulse-width, while
$t^i(n,m)$ denotes the delivery time of the $m$-th spike from the $n$-th inhibitory neuron 
to the $j$-th neuron. The sum $\sum^\prime_n$ is restricted to the $K^i$ neighbours of neuron $j$.
This representation amounts to assuming that $\mathcal{I}_j(t)$ is the linear superposition of 
the $\alpha$-pulses received by the neuron $j$ from inhbitory neurons until time $t$.
The excitatory field is treated in a slightly different way,
\begin{equation}
  \ddot{\mathcal{E}}^{e/i}_j + 2\alpha \dot{\mathcal{E}}^{e/i}_j + \alpha^2 \mathcal{E}^{e/i}_j = 
\alpha^2 \!\!\!\!\!\!\!\!\sum_{n, m|t^e(n,m)<t}^\prime
\!\!\!\!\!\!\! x^{{e/i}}_{n}  \delta(t-t^e(n,m))
\label{field_e}
\end{equation}
where $n$ identifies the excitatory neuron sending the $m$-th spike to the $j$-th neuron;
$x^{e/i} \in[0,1]$ represents the synaptic efficacy. If the receiving neuron is inhibitory $x^i \equiv 1$, 
while $x^e$ is affected by the STD acting on excitatory-to-excitatory connections.
By following \cite{tsodyks1998}, its evolution can be written as
\begin{equation}
	\dot x^e_n = \frac{(1-x^e_n)}{\tau_d}  -  u x^e_n \!\!\!\! \sum_{m|t^e(n,m)<t}\!\!\!\! \delta(t-t^e(n,m)) \; ,
\label{depr}
\end{equation}
where  $t^e(n,m)$ identifies the time of the $m$-th spike emitted by the  $n$-th neuron itself.
Whenever the neuron spikes, the synaptic efficacy $x^e_n$ is reduced by a factor $u$, 
representing the fraction of resources consumed to produce a post-synaptic spike.
So long as the $n$-th excitatory neuron does not spike, the variable $x^e_b$ 
increases towards 1 over a time scale $\tau_d$. 

The in-degrees of the two populations are distributed as in a massively coupled
Erd\"os-Renyi random graph~\cite{golomb2001}, i.e. setting  $K^{e/i} = p^{e/i} N$, 
where $p^{e/i} \in [0,1]$ is the probability to have a pre-synaptic connection.

\paragraph{Self-consistent approach}
Before discussing the direct numerical simulations, we present a simple self-consistent approach to explain
how STD can actually stabilize a balanced state even in absence of strong external currents.
In the above defined setup, $\mathcal{E}^{e/i}_j$ and $\mathcal{I}_j$, being proportional to the in-degree, are also proportional to $N$,
so that the two terms in Eq.~(\ref{eq:C}) are both proportional to $\sqrt{N}$. It is useful to make this dependence
explicit, by writing
\begin{equation}
C^{e/i}_j = [ \beta^{e/i}_e E^{e/i}_j - \beta^{e/i}_i I_j] \sqrt{N}  =  \Delta_j^{e/i} \sqrt{N}
\label{eq:c2}
\end{equation}
where $\beta^{e/i}_e = g^{e/i}_e \sqrt{p^e}$, $\beta^{e/i}_i = g^{e/i}_i \sqrt{p^i}$.
$I_j = \mathcal{I}_j/(p^iN)$ ($E^i_j = \mathcal{E}^i_j/(p^eN)$) represent the average firing rates of the
inhibitory (excitatory) pre-synaptic spike trains stimulating the $j$-th neuron; finally,
$E^e_j = \mathcal{E}^e_j/(p^eN)$ is the effective firing rate of an excitatory neuron scaled to account
for the reduced efficacy due to STD.
The approximation consists in neglecting neuron-to-neuron fluctuations, as well as temporal variations
so that we can drop the $j$ dependence of both the fields and the input currents and assume that they are
constant.
Within this approximation, a balanced regime can exist if $C^{e/i}$ remains finite
for $N\to\infty$, or, equivalently, if the terms in square brackets in Eq.~(\ref{eq:c2}) converge
to 0 (as $1/\sqrt{N}$).
Accordingly, 
\begin{equation}
E^i_\circ = \frac{\beta^i_i}{\beta^i_e} I_\circ \quad {\rm and} \quad E^e_\circ = \frac{\beta^e_i}{\beta^e_e}I_\circ \enskip .
\label{system}
\end{equation}
where the subscript ``$\circ$'' here and elsewhere means that the variable refers to the $N\to\infty$ limit.
Hence,
\begin{equation}
\frac{E^e_\circ}{E^i_\circ} = \frac{\beta_i^e\beta_e^i}{\beta^e_e\beta_i^i} =
\frac{g_i^eg_e^i}{g^e_eg_i^i} 
\label{eq:self0}
\end{equation}
In the absence of STD, since $E^e_\circ=E^i_\circ$, Eq.~(\ref{eq:self0}) is satified only when the rightmost h.s. is set equal to 1 from
the outset: hence, the balanced regime is highly non-generic.
In the literature, a way out is typically found by assuming that the external current $\omega$ is of order $\sqrt{N}$,
so that it must be included in the balance conditions (\ref{system}) as additional given terms.
As a result, the two equations compose a generically solvable, linear, inhomogeneous system~\cite{bal1}, the only condition
for its validity being that the fields must turn out positive.

In the present context, instead, the novelty is that the ratio between $E^i$ and $E^e$ is not
a priori equal to 1, but depends on the activity of the excitatory neurons.
In fact, $E^e = \theta E^i$, where $\theta$ is the value of the
synaptic efficacy when the excitatory neurons reach threshold during their periodic firing activity.

Hence, in the thermodynamic limit, the balance condition requires 
\begin{equation}
\theta_\circ = \frac{g_i^eg_e^i}{g^e_eg_i^i}  < 1
\label{eq:self1}
\end{equation}
where the inequality follows from the fact that $\theta_\circ$ must, by definition,
be smaller than 1.
In other words, a balanced regime is generic as it can arise for a broad range of coupling constants.
In this paper, since we set $g^e_e=1$, $g^e_i=1/2$, $g^i_e=1$, and $g^i_i=2$, the inequality is satisfied ($1/4<1$).

The equality between the first two terms in (\ref{eq:self1}) allows determining the value of
synaptic efficacy and, in turn, of the interspike interval $T^e_\circ$ of the excitatory neurons.
In fact, from the integration of Eq.~(\ref{depr}) in between two consecutive spikes,
\begin{equation}
\theta \equiv x^e(T^e) = 1 - (1-x^e(0))\mathrm{e}^{-T^e/\tau_d } \; .
\label{eqa:sd1-sol}
\end{equation}
The still unknown initial condition $x^e(0)$ can be determined by imposing
 $x^e(0) = u x^e(T^e)$ ($u$ is the depletion factor), obtaining
\begin{equation}
\theta = \frac{1-\mathrm{e}^{-T^e/\tau_d}}{1-u\mathrm{e}^{-T^e/\tau_d}} \; .
\label{eq:sd1b}
\end{equation}
By then solving for $T^e$, we find, in the thermodynamic limit, 
\begin{equation}
T^e_\circ = \ln \frac{1-u \theta_\circ}{1-\theta_\circ}
\label{eq:sd2}
\end{equation}
which means that the ISI and thereby the 
amplitude of the excitatory field
$E^i_\circ = 1/T^e_\circ$, are independent of the PRC (within this approximation).
Finally, the amplitude $I_\circ$ of the inhibitory field is obtained from the first of Eq.~(\ref{system}).

\paragraph{Mean Firing Rates}
The self-consistent analysis is useful to identify the necessary conditions for the onset
of a balanced regime, but it unavoidably predicts a current-driven regime.
In order to analyse the actual network behavior it is necessary to
perform numerical simulations. Here below, we report the results for
a homogenous network, where the bare firing rate is set to
$\omega = 50$ Hz and the PRC is $Z_I(\phi)$ 
(see~\cite{param} for the other parameter values).

In Fig.~\ref{fig:fig1}(b) we plot the population firing rates of excitatory 
$\nu^e = \langle E^i_j \rangle$ and inhibitory $\nu^i = \langle I_j \rangle$ 
neurons ($\langle \cdot \rangle$ denotes an average over all neurons of a given population) versus $N$. 
The data are well fitted by the law $\nu_0^{e/i} + \mu^{e/i}/\sqrt{N}$,
with $\nu_0^e\simeq \nu_0^i \simeq 5.78$ Hz, $\mu^e\simeq 399$ Hz, and 
$\mu^i\simeq 762$ Hz (curves not shown).
This indicates that the single-neuron activity remains finite for $N\to \infty$, a
clear signature of a balanced regime. This conclusion is confirmed by the $N$-dependence of the
average unbalance $\Delta^{e/i} = \langle \Delta_j^{e/i} \rangle$,
reported in Fig.~\ref{fig:fig1}(c), where a clear $1/\sqrt{N}$ decrease is visible,
implying that the average value of $C^{e/i}$ stays constant for $N\to\infty$.

It is instructive to compare the numerical results with the semi-analytical perturbative
implementation of the self-consistent approximation.
In the previous section, we have shown how to determine the 
values of the excitatory and inhibitory fields for $N \to \infty$.
For our parameter values, $I_\circ = E^i_\circ=1/\log(7/6) {\rm Hz} \simeq 6.487$ Hz ~\cite{note1}.
In the Supplementary Material, we show how to go beyond the asymptotic values,
determining finite-size corrections. Here, we sketch the procedure.
From the knowledge of the asymptotic fields, one can determine the 
input currents $C^{e/i}_\circ$ responsible for those fields,
by integrating Eq.~(\ref{eq:model0}) under the assumption of a constant current. Then, the definition 
(\ref{eq:c2}) of $C^{e/i}$ can be used as a consistency relationship to 
determine the finite-size corrections for both fields, which turn out to be
proportional to $1/\sqrt{N}$. The resulting analytic expressions are reported in the Supplementary Material
and plotted in Fig.~\ref{fig:fig1}(b) (see the solid curves). They 
overestimate the numerical values, but are not too far from them.
 
\begin{figure}
\begin{centering}
\includegraphics[width=0.45\textwidth,clip=true]{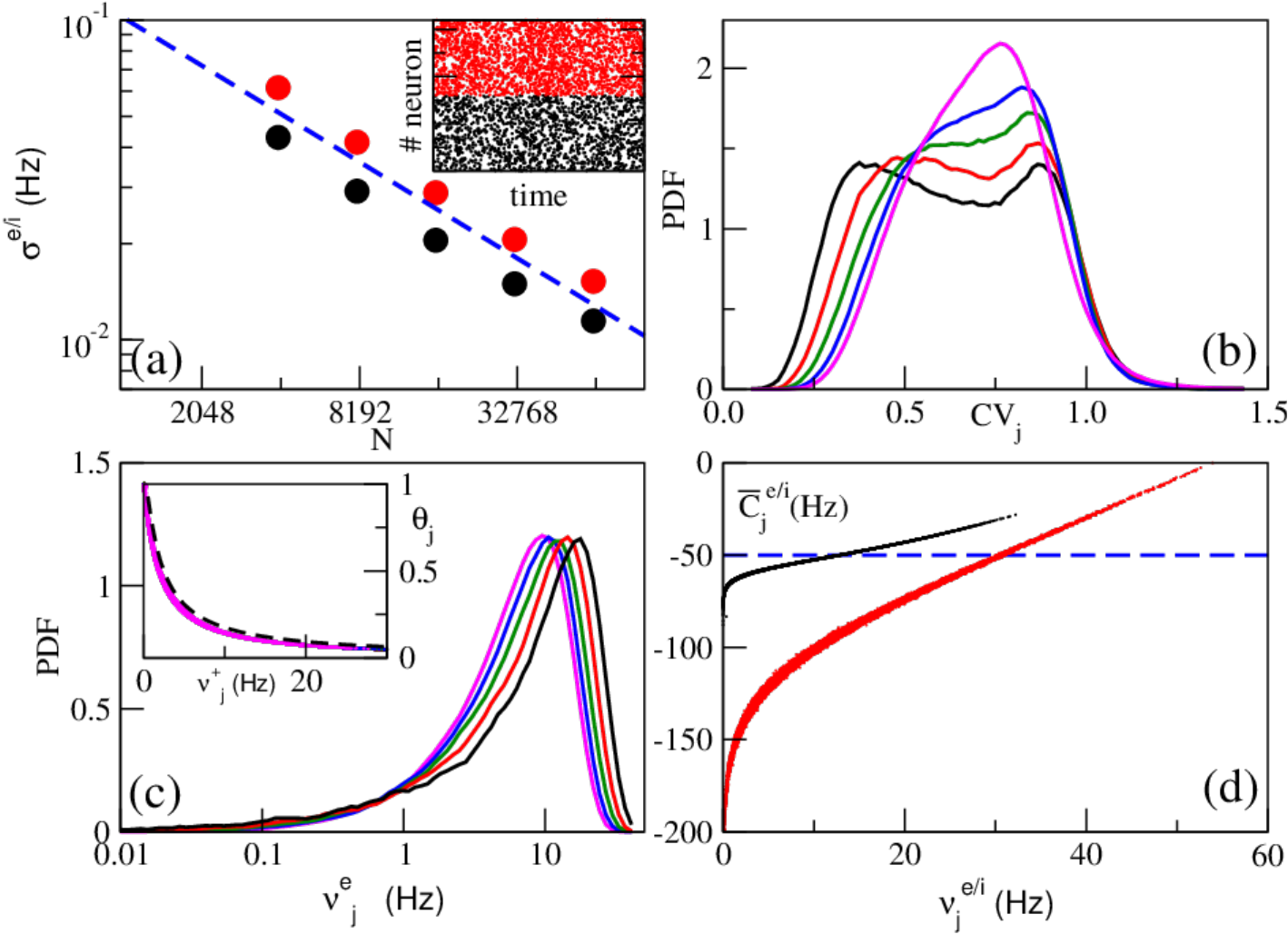}
\caption{(a) Standard deviation $\sigma^{{e/i}}$ of the 
population firing rates versus $N$, the blue dashed line indicates a power law decay as $1/\sqrt{N}$.
The inset displays a raster plot for $N=16000$ over a time window of 20 ms.
PDF of the coefficients of variation CV$_i$ (b) and of the firing rates 
$\nu_j^e$ (c) for excitatory neurons.
In the inset of (c), the synaptic efficacy at threshold $\theta_j$ is displayed versus the corresponding firing rate $\nu_j^e$: the dashed line
is the self-consistent estimation \eqref{estimate}. (d) 
Time averaged values ${\overline C}^e_j$ (${\overline C}^i_j$) versus the corresponding firing rates $\nu_j^e$ ($\nu_j^i$) for $N=16000$. The dashed line denotes the value  $C^{e/i} = -50$ Hz discriminating fluctuation from current driven dynamics.
The black (red) color in (a-d) refers to excitatory (inhibitory) neurons, while
the colors in (b-c) to different system sizes : namely, $N=8000$ (black);
$N=16000$ (red); $N=32000$ (green); $N=64000$ (blue) and $N=128000$ (magenta). The reported
data refer to $Z_{\rm I}$.
}
\label{fig:fig2}
\end{centering}
\end{figure}

\paragraph{Fluctuations}
We start investigating whether the collective dynamics of the network
remains asynchronous even for large system sizes, as expected in 
brain circuits \cite{bal2}. 
The raster plot for $N=16000$, reported in the inset of Fig.~\ref{fig:fig2}(a),
does not reveal any population oscillation.
A more quantitative analysis has been made by computing the time-average of the standard deviation 
of the incoming fields (i.e. of the instantaneous population firing rates), here denoted with $\sigma^{e/i}$.
The values computed for different network sizes, reported in Fig.~\ref{fig:fig2}(a),
decrease consistently with the $1/\sqrt{N}$ scaling expected 
from the central limit theorem for an  asynchronous dynamics.

Next, we focus on temporal fluctuations at the single-neuron level. They are disregarded a priori 
by the self-consistent approach, but the probability distribution density (PDF) of the coefficient of 
variations CV$_j$ reported in Fig.~\ref{fig:fig2}(b) for the excitatory neurons gives a clear
evidence of irregularity. Some neurons are characterized by a CV even larger than 1, the value
expected for a Poisson distribution, and the irregularity tends to increase with $N$.
A similar scenario is exhibited by inhibitory neurons (data not shown).

Finally, we turn our attention to ensemble fluctuations.
The firing rates themselves are broadly distributed from nearly vanishing values (almost silent neurons) up to 50-60 Hz, with a pronounced peak around 5-10 Hz.
When $N$ is increased, the PDF widths remain finite and their shapes appear to converge to a given asymptotic form,
as clearly visible in Fig.~\ref{fig:fig2}(c) where the data refer to excitatory neurons. 
This manifestation of heterogeneity is not surprising in a massively coupled Erd\"os-Renyi network. In fact, the single-neuron connectivity is expected to exhibit fluctuations of order $\sqrt{N}$, which transform themselves into fluctuations of ${\cal O}(1)$ for the $C_j^{e/i}$, and therefore for the firing rates. 

The distribution of firing rates $\nu^e_j$ induces a distribution of synaptic efficacies $\theta_j$
(taken in correspondence of the spiking times). 
Under the approximation of negligible temporal fluctuations, one can reformulate Eq.~(\ref{eq:sd1b}) as
\begin{equation}
 \theta_j  = \frac{1-\rm{e}^{-1/(\nu^e_j \tau_d)}}{1- u \rm{e}^{-1/(\nu^e_j \tau_d)}} \; .
\label{estimate}
\end{equation}
The inset in Fig.~\ref{fig:fig2}(c) reveals a good agreement with the numerical simulations.

The PDF shapes reported in Fig.~\ref{fig:fig2}(c), are similar to those
measured experimentally in the cortex and hippocampus~\cite{hromadka2008,o2010,wohrer2013, buzsaki2014,mongillo2018},
with many neurons exhibiting a low firing rate and a high-frequency long tail, akin to  
a log-normal distribution. 
These shapes are typically interpreted as an indication of
fluctuation-driven dynamics~\cite{roxin2011}. 
It is therefore convenient to test whether the neurons, in our model, operate either above or below threshold.
This can be done as follows.
From Eq.~(\ref{eq:model0}), since the maximum of $Z(v^{e/i})$ is 1,
$\dot v^{e/i}$ may have a stable zero, only if $\omega + GC^{e/i}<0$.
Hence,
a neuron characterized by an average current ${\overline C}^{e/i}$ is typically  below
threshold if ${\overline C}^{e/i} < -\omega /G= -50$ Hz.
The data reported in Fig.~\ref{fig:fig2}(d) reveal a  mixed behavior: depending on their effective
firing rate, neurons may be either fluctuation- or current-driven. 
By further averaging over the entire population, we
find that while the inhibitory neurons are significantly fluctuation-driven with $\langle {\overline C}^i \rangle = -101.5$ Hz, 
on average the excitatory neurons operate slightly below threshold being $\langle {\overline C}^e \rangle = -55.5$ Hz.
Altogether, the network self-stabilizes in a regime, where fluctuations play a major role, 
consistently with the observation of a pseudo log-normal distribution of the firing rates \cite{roxin2011}.

\begin{figure}
\begin{centering}
\includegraphics[width=0.35\textwidth,clip=true]{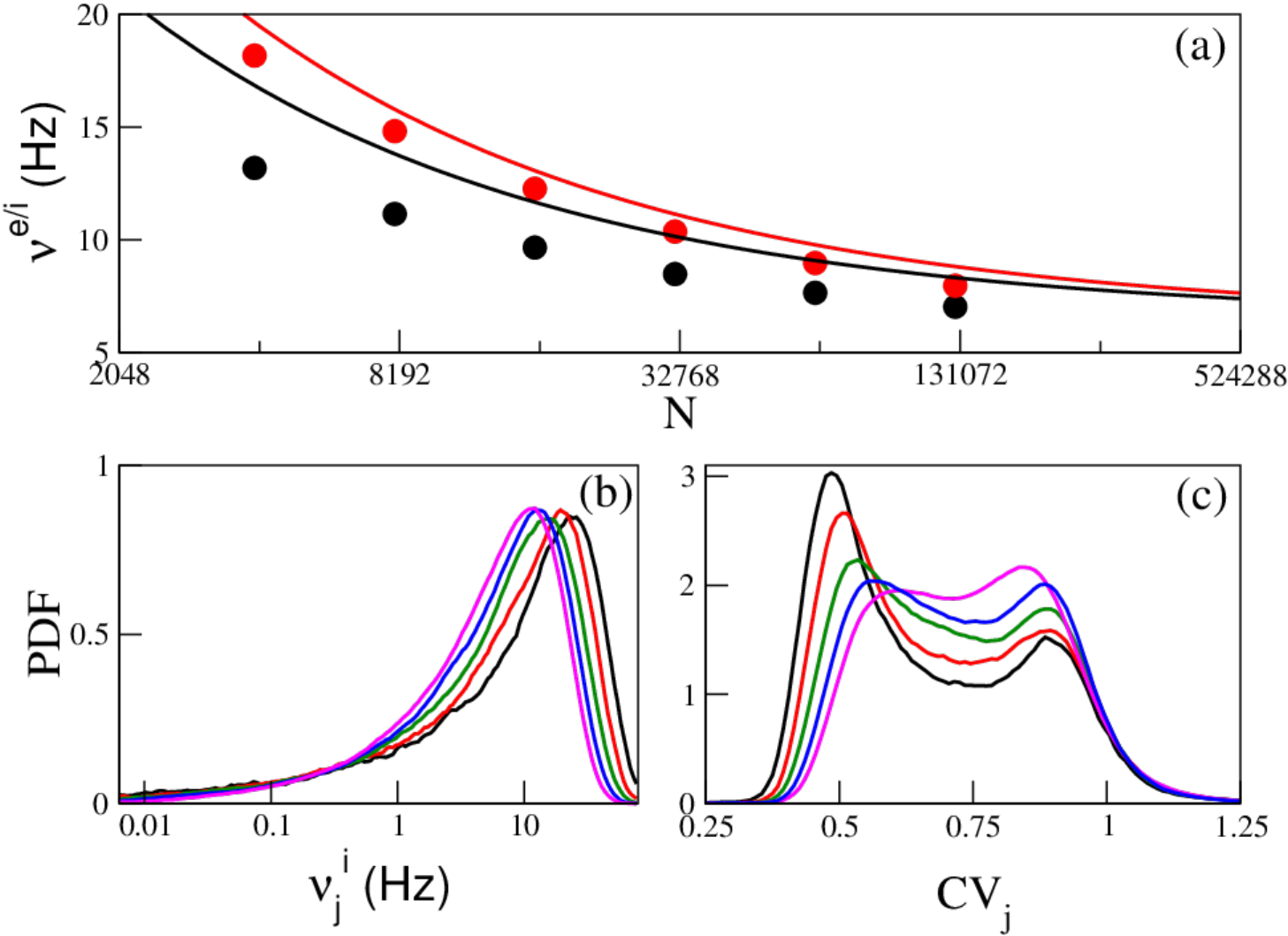}
\caption{(a) The average population firing rates $\nu^e$ (black circles) and 
$\nu^i$  (red circles) versus $N$; PDF of the inhibitory firing rates $\nu^i_j$ (b)
and of the corresponding CV$_j$ (c).
In (a) the black (red) solid line indicates the self-consistent approximation for
excitatory (inhibitory) neurons and the colors in (b-c) different system sizes 
coded as in Fig. \ref{fig:fig2} (b-c). The results here reported refer to the LIF model.
}
\label{fig:fig3}
\end{centering}
\end{figure}

\paragraph{Robustness of the mechanism}
Additional simulations performed for different parameter values and even introducing heterogeneity in the external currents
have shown the generality of the mechanism. Details can be found in the Supplementary Material.
Here we focus on the most important test, made by using the PRC of LIF neurons, $Z_{\rm LIF}$.  All parameters have been left unchanged, except for a faster synaptic transmission $\alpha$~\cite{param} to check the specificity of the pulsewidth.
The simulations are performed by integrating Eq. \eqref{eq:model0} with $Z_{\rm LIF}$.
As shown analytically (and verified numerically) this model is fully equivalent to 
a standard current-driven spiking LIF network \eqref{eq:model00} with $a=e/(e-1) \simeq 1.582$ and $G^\prime=1/(e-1) \simeq 0.582$.

In~Fig.~\ref{fig:fig3} (a) we report the firing rates of the two populations (black and red dots), together with the outcome of
the self-consistent approach (solid curves). 
The theoretical curves converge, as they should, to the same value $1/T_\circ$, which coincides with the previous 
asymptotic value since, from its definition, it does not depend on the PRC shape.
Also the numerically determined firing rates converge to the same value, again following the law
$\nu^{e/i} \simeq \nu_0 + \mu^{e/i}/\sqrt{N}$, where
$\nu_0 \simeq 5.72$ Hz, $\mu^e \simeq 480$ Hz and, and $\mu^i=803$ Hz (curves not shown).
The smallness of the discrepancy with the previous asymptotic value ($\simeq 1\%$) indicates 
that the irrelevance of the PRC extends to the complete model, where all kinds of fluctuations are automatically
included. 

In Fig.~\ref{fig:fig3} we also report the PDF of the firing rates of inhibitory neurons
$\nu^i_j$ (panel b) and of the corresponding coefficients of variations CV$_j$  (panel c) for different network sizes. The distributions of the firing rates display a long tail towards vanishing rates, a typical feature of neurons which operate below threshold and that are therefore fluctuation driven, as confirmed by the values of the corresponding CV$_j$. 
Analogous results have been found for the excitatory neurons (data not shown).
We can safely state that the data obtained for the LIF network confirm the scenario
of a balanced regime as for the $Z_{\rm I}(\phi)$ PRC.

\paragraph{Conclusions}

The typical setup studied in the literature to discuss dynamically balanced regimes requires 
the presence of strong external inputs~\cite{bal1}. 
An alternative layout, which does not suffer this limitation, was proposed in Ref.~\cite{khajeh2022}, together
with the concept of {\it sparse balance}. It, however,
requires an anomalously broad distribution of synaptic
strengths and leads to a vanishing fraction of active neurons (in the thermodynamic limit).

The mechanism proposed here is more robust and general: 
it exploits the dynamical adjustment of the synaptic currents, resulting from
short-term synaptic depression (STD). 
STD is a much studied mechanism, already invoked to explain fundamental cognitive functions, such as
working memory~\cite{del2003,mongillo2008,taher2020} and the internal representation of spatio-temporal 
information~\cite{romani2015, wang2015, haimerl2019, pietras2022}.
By lowering the strength of highly active excitatory connections, STD eventually binds the activity of excitatory neurons.

Mathematically, the balanced regime is made possible by the nonlinear dependence of one excitatory 
current on the amplitude of the corresponding field. In practice, the homogeneous set of linear conditions
\eqref{system}, which emerges for weak external currents, is transformed into a nonlinear one.
While the former one admits a meaningful solution only for a special combination of the coupling constant,
the latter is generically solvable for a broad range of parameter values.
Once this has been understood, it is straightforward to infer that other nonlinear mechanisms
may play the same role as STD in the absence of strong inputs.
In fact, other sources of naturally expected nonlinearities have been recently investigated in computational neuroscience
although, always in the presence of strong external currents.
Spike-frequency adaptation is one such  mechanism, studied in networks with highly heterogenous in-degrees~\cite{landau2016}.
Similarly, facilitation has been found to promote the emergence of bistable balanced regimes~\cite{mongillo2012,hansel2013}.
It is time to move on and to investigate all such mechanisms in the absence of strong external currents.

\section*{Supplementary Material}

\subsection*{Self-Consistent Analysis}

Here we develop a perturbative analysis to determine the mean field properties of the asynchronous regime, 
for a homogeneous setup, via a self-consistent approach which neglects neuron-to-neuron fluctuations.
Within this approximation, the neuron label $j$ can be dropped into the corresponding evolution equations
\begin{equation}
        \dot \phi^{e/i} = \omega + GZ(\phi^{e/i}) C^{e/i} \enskip ;
\label{eq:model0}
\end{equation}
where
\begin{equation}
C^{e/i} = [ \beta^{e/i}_e E^{e/i} - \beta^{e/i}_i I] \sqrt{N}  \enskip ;
\label{eq:c2}
\end{equation}
and the field $E^{e/i}$ ($I$) represents the mean firing rate of the pre-synaptic excitatory (inhibitory) neurons, which, for an asynchronous regime, is constant. 
In this case the firing period is
\begin{equation}
    T^{e/i} = \int_0^1 \frac{d \phi}{\omega + C^{e/i} G Z(\phi)} \enskip .
    \label{period}
\end{equation}

In the paper, we have shown how to determine the asymptotic values (i.e. for $N\to\infty$) of all the fields.
By integrating Eq.~(\ref{eq:model0}), we can now determine $C_\circ^{e/i}$ as as those values which yield the expected 
firing period $T_\circ^{e/i}$ (i.e. the expected field).
For the PRC $Z_{\rm I}(\phi)$ we find $C^e_\circ=C^i_\circ=-49.108$ Hz, i.e. both families of neurons operate 
slightly above threshold since we set $\omega = 50$ Hz.

For the LIF model we simply have
\begin{equation}
    T_\circ^{e/i} = \int_0^1 \frac{d \phi}{\omega + C^{e/i}_\circ G \mathrm{e}^{\phi-1}} =
    \frac{Y^{e/i}}{\omega} \int_1^\mathrm{e} \frac{dv}{v(Y^{e/i}+v)}  \enskip ,
\end{equation}
where $v = \exp{\phi}$ and $Y^{e/i} = (\omega \mathrm{e})/(C^{e/i}_\circ G)$ ($Y^{e/i} <-\mathrm{e}$)
Hence,
\begin{equation}
    T^{e/i}_\circ = 
    \frac{1}{\omega} \left[1-\ln \frac{Y^{e/i}+\mathrm{e}}{Y^{e/i}+1} \right] \quad .
    \label{tinf_LIF}
    \end{equation}

For the values selected in our paper~\cite{param}, $T^{e/i}_\circ = \ln 7/6  = 0.15415$ s, so that 
$I_\circ = E^i_\circ = 1/T^e_\circ = 6.487$ Hz and these values do not depend on the chosen PRC.
By inserting our parameter values, we find that $Y^{e/i}=-2.7204$, which corresponds to $C^{e/i}_\circ=-49.96$ Hz.
Hence, in the LIF case,
the neurons families operate (in this mean field approximation) even closer to threshold.

Now, we can focus on Eq.~(\ref{eq:c2}) to obtain the first-order corrections to the fields since $C^e_\circ$ and $C^i_\circ$ are known,
\begin{eqnarray}
&&\beta^i_e E^i - \beta^i_i I = \frac{C^i_\circ}{\sqrt{N}} \enskip , 
\label{e1}\\
&& \beta^e_e \theta E^i  - \beta^e_iI = \frac{C^e_\circ}{\sqrt{N}}  \enskip 
\label{e2} ;
\end{eqnarray}
where we have exploited the equality $E^e = \theta E^i$, $\theta$ being the value of the synaptic efficacy when the excitatory neurons reach threshold
during their periodic firing activity. The value of $\theta$ can be determined from Eq.~(12) in the main text, here rewritten as
\begin{equation}
\theta=\frac{1-{\rm e}^{-\mathcal{T}}}{1-u{\rm e}^{-\mathcal{T}}}
\label{teta}
\end{equation}
with $\mathcal{T} = T^e/\tau_d$, to simplify the notations.

By solving the equation \eqref{e1} for $I$
\[
 I = \frac{\beta^i_e}{\beta^i_i}E^i  - \frac{C_\circ^i}{\beta^i_i\sqrt{N}} \enskip,
\]
 and  replacing it in \eqref{e2},
\[
[\beta^i_i\beta^e_e \theta - \beta^e_i\beta^i_e] E^i 
 = \frac{\beta^i_iC_\circ^e - \beta^e_i C_\circ^i }{\sqrt{N}}  \enskip .
\]

By employing the explicit expression \eqref{teta} for $\theta$, the above equation can be rewritten as
\[
\frac{1}{\mathcal{T}} 
 \left [ 
\beta^i_i\beta^e_e 
\frac{1- \mathrm{e}^{- \mathcal{T}}}{1-u \mathrm{e}^{- \mathcal{T}}}  - \beta^e_i\beta^i_e 
\right ]
 = \frac{\beta^i_iC^e_\circ - \beta^e_i C^i_\circ }{\sqrt{N}} \tau_d \enskip ,
\]
and hence,

\[
\frac{\beta^i_i\beta^e_e}{\mathcal{T}} 
 \left [ 
\frac{
(1 - \theta_\circ) \mathrm{e}^\mathcal{T}- 1 +
 u \theta_\circ}
{\mathrm{e}^\mathcal{T}-u}
\right ]
 = \frac{\beta^i_iC^e_\circ - \beta^e_i C^i_\circ }{\sqrt{N}} \tau_d \enskip ;
\]
where $\theta_\circ= \frac{\beta^e_i\beta^i_e}{\beta^i_i\beta^e_e} = \frac{g^e_ig^i_e}{g^i_ig^e_e}$ 
is the asymptotic ($N \to \infty$) $\theta$ value as defined in Eq. (10) in the main text.

Now, we expand 
\[
\mathrm{e}^\mathcal{T} = \mathrm{e}^{T^e_\circ/\tau_d} + \delta  = \frac{1-u\theta_\circ}{1-\theta_\circ} + \delta
\]
At first order, it is sufficent to expand the numerator of the l.h.s.

\[
\delta
 = \frac{\beta^i_iC^e_\circ - \beta^e_i C^i_\circ }{\beta^i_i\beta^e_e } \frac{1-u}{(1-\theta_\circ)^2} \frac{T^e_\circ}{\sqrt{N}}
\]

\[
T^e = \mathcal{T}\tau_d =  T^e_\circ + \frac{1-u\theta_\circ}{1-\theta_\circ} \delta \tau_d
\]
and finally
\[
E^i = \frac{1}{T^e_\circ} \left [  1 - 
 \frac{\beta^i_iC^e_\circ - \beta^e_i C^i_\circ }{\beta^i_i\beta^e_e } 
\frac{(1-u)}{(1-\theta_\circ)(1-u\theta_\circ)} \frac{\tau_d}{\sqrt{N}} \right ]
\]
The perturbative expression for $I$ is
\[
 I = \frac{1}{T^e_\circ} \frac{\beta^i_e}{\beta^i_i}   - \left [
 \frac{1}{T_\circ} \frac{\beta^i_e}{\beta^i_i}
 \frac{\beta^i_iC^e_\circ - \beta^e_i C^i_\circ }{\beta^i_i\beta^e_e } 
\frac{(1-u)\tau_d}{(1-\theta_\circ)(1-u \theta_\circ)} + \frac{C^i_\circ}{\beta^i_i}\right ] \frac{1}{\sqrt{N}} 
\]
In our setup \cite{param}
\begin{equation}
E^i = \frac{1}{T_\circ} \left [ 1 -
 \sqrt{2} 
\frac{10}{7}
\frac{C_\circ}{\sqrt{N}} \right]
\label{exc_corr}
\end{equation}
and
\begin{equation}
 I = \frac{1}{T_\circ}   - \left [
 \frac{1}{7T_\circ}
 + \frac{1}{4}
\right ] 
\frac{10C_\circ \sqrt{2}}{\sqrt{N}} 
\label{inh_corr}
\end{equation}
where we have set $C^{e/i}_\circ=C_\circ$ and $T^{e/i}_\circ=T_\circ$, since they coincide for our
choice of the parameters.

For $Z_{\rm I}(\phi)$ we have found numerically $T_\circ$ and $C_\circ$ and therefore
also the values for the following perturbative expansion to the first order:
\[
E^i = 6.487  + 643.61/\sqrt{N} \qquad , \qquad
I = 6.487+817.23/\sqrt{N} \qquad , \qquad
\]
where $E^i$ and $I$ are expressed in Hz.

For the LIF model we have instead found analytically that :
\[
E^i = 6.487  + 654.76/\sqrt{N} \qquad , \qquad
I = 6.487 + 831.40/\sqrt{N} \qquad , \qquad
\]
where the first order corrections are slightly larger than in the case of $Z_{\rm I}(\phi)$.

\subsection*{Heterogeneous Network and Further Dynamical Regimes}

\begin{figure}
\begin{centering}
\includegraphics[width=0.45\textwidth,clip=true]{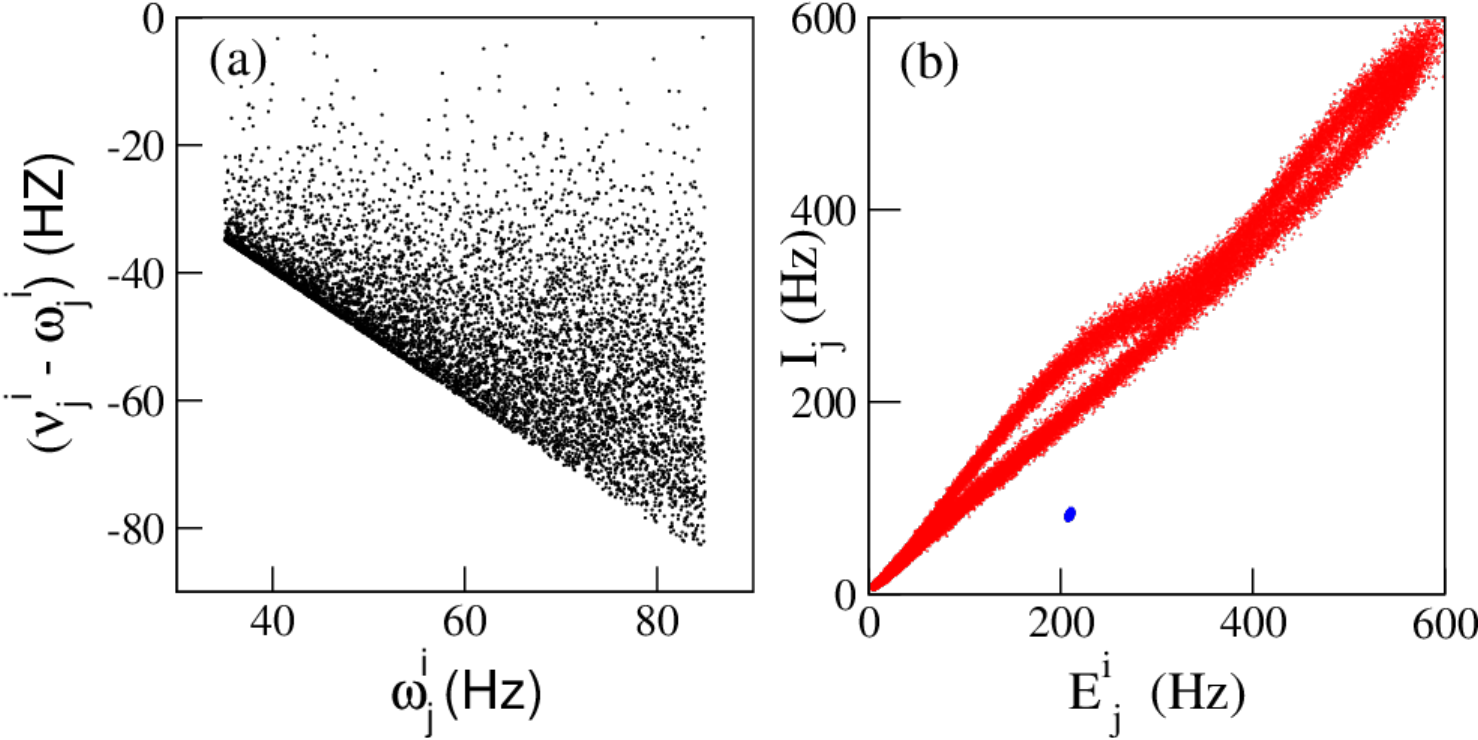}
\caption{(a) Difference between the single neuron firing rates and their bare fequencies $\nu_j^i - \omega_j^i$ 
versus $\omega_j^i$ for an heterogeneous network with $N=16000$ neurons with  $\omega^e_j$ ($\omega^i_j$) 
	uniformly distributed in the interval $[15,65]$ Hz ($[35,85]$ Hz).
(b) Inhibitory fields $I_j$ versus the corresponding excitatory field $E^i_j$ for a homogeneous 
	network with $N=16000$ and $\tau_d = 1$ s (blue dots) and $\tau_d= 0.1$ s (red dots). 
	The values of the fields for $\tau_d= 1$ s ar arbitrarily shifted for a better visualization.
	The reported results refer to the PRC $Z_{\rm I}(\phi)$.
}
\label{fig:fig4}
\end{centering}
\end{figure}

In the paper, we have investigated a homogeneous setup, where all neurons are characterized by the same bare frequency,
but this is definitely an idealization of reality. In order to further validate the robustness of the proposed
mechanism, we have also studied a  model where the bare frequencies are uniformly distributed over two different
ranges (for the excitatory and inhibitory neurons). A balanced regime emerges again, where the network-induced
heterogeneity adds up to the spontaneous single-neuron heterogeneity. In Fig.~\ref{fig:fig4}(a),
we plot the difference between the effective $\nu_j^i$ and the bare frequency $\omega_j^i$, versus 
the bare frequency itself (for each
inhibitory neuron) to highlight the impact of the network coupling. An overall inhibitory effect is
noticeable, which increases upon increasing the bare frequency.

We have also performed various simulations for different sets of parameter values, confirming that an
asynchronous balanced state is a generic regime in networks accompanied by STD.
The only qualitative change we have found is the onset of collective oscillations,
typically when some time scales are varied. As an example, 
in Fig.\ref{fig:fig4}(b) one can see relatively wide oscillations of both the excitatory and inhibitory fields
(see the red dots), to be confronted with the small blue spot which corresponds to the asynchronous regime studied
in detail in this Letter.
The structure of the attractor is suggestive of an underlying possibly low-dimensional dynamics, but a simulation
performed for a single (actually a few) networks sizes does not suffice to conclude whether the fatness of the
attractor is either a finite-size effect, or the manifestation of an intrinsic high-dimensionality.

\section*{acknowledgments}
We thank German Mato for useful discussions in the initial stage of this project, during 
our stay at Max Planck Institute for the Physics of Complex Systems
(Dresden, Germany) within the Advanced
Study Group “From Microscopic to Collective Dynamics
in Neural Circuits” (2016/17).
A.T. also  acknowleges useful interactions with Gianluigi Mongillo and Nina La Miciotta. A. T. received financial support
by the ANR Project ERMUNDY (Grant No. ANR-18-CE37-0014), by the Labex MME-DII (No. ANR-11-
LBX-0023-01) and by CY Generations (Grant No ANR-21-EXES-0008), all part of the  French programme “Investissements d’Avenir”. A.P. received support by CY Advanced Studies (Cergy-Pontoise, France) for a visiting scholarship in 2018.

 \section*{DATA AVAILABILITY}

The data that support the findings of this study are available on request from the authors .

  
\bibliographystyle{apsrev4-1}  
\bibliography{lif_ost_cd}

\begin{thebibliography}{71}%
\makeatletter
\providecommand \@ifxundefined [1]{%
 \@ifx{#1\undefined}
}%
\providecommand \@ifnum [1]{%
 \ifnum #1\expandafter \@firstoftwo
 \else \expandafter \@secondoftwo
 \fi
}%
\providecommand \@ifx [1]{%
 \ifx #1\expandafter \@firstoftwo
 \else \expandafter \@secondoftwo
 \fi
}%
\providecommand \natexlab [1]{#1}%
\providecommand \enquote  [1]{``#1''}%
\providecommand \bibnamefont  [1]{#1}%
\providecommand \bibfnamefont [1]{#1}%
\providecommand \citenamefont [1]{#1}%
\providecommand \href@noop [0]{\@secondoftwo}%
\providecommand \href [0]{\begingroup \@sanitize@url \@href}%
\providecommand \@href[1]{\@@startlink{#1}\@@href}%
\providecommand \@@href[1]{\endgroup#1\@@endlink}%
\providecommand \@sanitize@url [0]{\catcode `\\12\catcode `\$12\catcode
  `\&12\catcode `\#12\catcode `\^12\catcode `\_12\catcode `\%12\relax}%
\providecommand \@@startlink[1]{}%
\providecommand \@@endlink[0]{}%
\providecommand \url  [0]{\begingroup\@sanitize@url \@url }%
\providecommand \@url [1]{\endgroup\@href {#1}{\urlprefix }}%
\providecommand \urlprefix  [0]{URL }%
\providecommand \Eprint [0]{\href }%
\providecommand \doibase [0]{http://dx.doi.org/}%
\providecommand \selectlanguage [0]{\@gobble}%
\providecommand \bibinfo  [0]{\@secondoftwo}%
\providecommand \bibfield  [0]{\@secondoftwo}%
\providecommand \translation [1]{[#1]}%
\providecommand \BibitemOpen [0]{}%
\providecommand \bibitemStop [0]{}%
\providecommand \bibitemNoStop [0]{.\EOS\space}%
\providecommand \EOS [0]{\spacefactor3000\relax}%
\providecommand \BibitemShut  [1]{\csname bibitem#1\endcsname}%
\let\auto@bib@innerbib\@empty
\bibitem [{\citenamefont {van Vreeswijk}\ and\ \citenamefont
  {Sompolinsky}(1996)}]{bal1}%
  \BibitemOpen
  \bibfield  {author} {\bibinfo {author} {\bibfnamefont {C.}~\bibnamefont {van
  Vreeswijk}}\ and\ \bibinfo {author} {\bibfnamefont {H.}~\bibnamefont
  {Sompolinsky}},\ }\href {\doibase 10.1126/science.274.5293.1724} {\bibfield
  {journal} {\bibinfo  {journal} {Science}\ }\textbf {\bibinfo {volume}
  {274}},\ \bibinfo {pages} {1724} (\bibinfo {year} {1996})}\BibitemShut
  {NoStop}%
\bibitem [{\citenamefont {Kuramoto}(2012)}]{kuramoto2012}%
  \BibitemOpen
  \bibfield  {author} {\bibinfo {author} {\bibfnamefont {Y.}~\bibnamefont
  {Kuramoto}},\ }\href@noop {} {\emph {\bibinfo {title} {Chemical oscillations,
  waves, and turbulence}}},\ Vol.~\bibinfo {volume} {19}\ (\bibinfo
  {publisher} {Springer Science \& Business Media},\ \bibinfo {year}
  {2012})\BibitemShut {NoStop}%
\bibitem [{\citenamefont {Acebr{\'o}n}\ \emph {et~al.}(2005)\citenamefont
  {Acebr{\'o}n}, \citenamefont {Bonilla}, \citenamefont {Vicente},
  \citenamefont {Ritort},\ and\ \citenamefont {Spigler}}]{acebron2005}%
  \BibitemOpen
  \bibfield  {author} {\bibinfo {author} {\bibfnamefont {J.~A.}\ \bibnamefont
  {Acebr{\'o}n}}, \bibinfo {author} {\bibfnamefont {L.~L.}\ \bibnamefont
  {Bonilla}}, \bibinfo {author} {\bibfnamefont {C.~J.~P.}\ \bibnamefont
  {Vicente}}, \bibinfo {author} {\bibfnamefont {F.}~\bibnamefont {Ritort}}, \
  and\ \bibinfo {author} {\bibfnamefont {R.}~\bibnamefont {Spigler}},\
  }\href@noop {} {\bibfield  {journal} {\bibinfo  {journal} {Reviews of modern
  physics}\ }\textbf {\bibinfo {volume} {77}},\ \bibinfo {pages} {137}
  (\bibinfo {year} {2005})}\BibitemShut {NoStop}%
\bibitem [{\citenamefont {Sherrington}\ and\ \citenamefont
  {Kirkpatrick}(1975)}]{sherrington1975}%
  \BibitemOpen
  \bibfield  {author} {\bibinfo {author} {\bibfnamefont {D.}~\bibnamefont
  {Sherrington}}\ and\ \bibinfo {author} {\bibfnamefont {S.}~\bibnamefont
  {Kirkpatrick}},\ }\href@noop {} {\bibfield  {journal} {\bibinfo  {journal}
  {Physical review letters}\ }\textbf {\bibinfo {volume} {35}},\ \bibinfo
  {pages} {1792} (\bibinfo {year} {1975})}\BibitemShut {NoStop}%
\bibitem [{\citenamefont {Kirkpatrick}\ and\ \citenamefont
  {Sherrington}(1978)}]{kirkpatrick1978}%
  \BibitemOpen
  \bibfield  {author} {\bibinfo {author} {\bibfnamefont {S.}~\bibnamefont
  {Kirkpatrick}}\ and\ \bibinfo {author} {\bibfnamefont {D.}~\bibnamefont
  {Sherrington}},\ }\href@noop {} {\bibfield  {journal} {\bibinfo  {journal}
  {Physical Review B}\ }\textbf {\bibinfo {volume} {17}},\ \bibinfo {pages}
  {4384} (\bibinfo {year} {1978})}\BibitemShut {NoStop}%
\bibitem [{\citenamefont {Antoni}\ and\ \citenamefont
  {Ruffo}(1995)}]{antoni1995}%
  \BibitemOpen
  \bibfield  {author} {\bibinfo {author} {\bibfnamefont {M.}~\bibnamefont
  {Antoni}}\ and\ \bibinfo {author} {\bibfnamefont {S.}~\bibnamefont {Ruffo}},\
  }\href@noop {} {\bibfield  {journal} {\bibinfo  {journal} {Physical Review
  E}\ }\textbf {\bibinfo {volume} {52}},\ \bibinfo {pages} {2361} (\bibinfo
  {year} {1995})}\BibitemShut {NoStop}%
\bibitem [{\citenamefont {Ginelli}\ \emph {et~al.}(2011)\citenamefont
  {Ginelli}, \citenamefont {Takeuchi}, \citenamefont {Chat{\'e}}, \citenamefont
  {Politi},\ and\ \citenamefont {Torcini}}]{ginelli2011}%
  \BibitemOpen
  \bibfield  {author} {\bibinfo {author} {\bibfnamefont {F.}~\bibnamefont
  {Ginelli}}, \bibinfo {author} {\bibfnamefont {K.~A.}\ \bibnamefont
  {Takeuchi}}, \bibinfo {author} {\bibfnamefont {H.}~\bibnamefont {Chat{\'e}}},
  \bibinfo {author} {\bibfnamefont {A.}~\bibnamefont {Politi}}, \ and\ \bibinfo
  {author} {\bibfnamefont {A.}~\bibnamefont {Torcini}},\ }\href@noop {}
  {\bibfield  {journal} {\bibinfo  {journal} {Physical Review E}\ }\textbf
  {\bibinfo {volume} {84}},\ \bibinfo {pages} {066211} (\bibinfo {year}
  {2011})}\BibitemShut {NoStop}%
\bibitem [{\citenamefont {Konishi}\ and\ \citenamefont
  {Kaneko}(1990)}]{konishi1990}%
  \BibitemOpen
  \bibfield  {author} {\bibinfo {author} {\bibfnamefont {T.}~\bibnamefont
  {Konishi}}\ and\ \bibinfo {author} {\bibfnamefont {K.}~\bibnamefont
  {Kaneko}},\ }\href@noop {} {\bibfield  {journal} {\bibinfo  {journal}
  {Journal of Physics A: Mathematical and General}\ }\textbf {\bibinfo {volume}
  {23}},\ \bibinfo {pages} {L715} (\bibinfo {year} {1990})}\BibitemShut
  {NoStop}%
\bibitem [{\citenamefont {Winfree}(2001)}]{winfree}%
  \BibitemOpen
  \bibfield  {author} {\bibinfo {author} {\bibfnamefont {A.~T.}\ \bibnamefont
  {Winfree}},\ }\href@noop {} {\emph {\bibinfo {title} {The Geometry of
  Biological Time}}},\ \bibinfo {edition} {2nd}\ ed.,\ \bibinfo {series}
  {Interdisciplinary Applied Mathematics}, Vol.~\bibinfo {volume} {12}\
  (\bibinfo  {publisher} {Springer-Verlag New York},\ \bibinfo {year}
  {2001})\BibitemShut {NoStop}%
\bibitem [{\citenamefont {Strogatz}\ and\ \citenamefont
  {Stewart}(1993)}]{strogatz1993}%
  \BibitemOpen
  \bibfield  {author} {\bibinfo {author} {\bibfnamefont {S.~H.}\ \bibnamefont
  {Strogatz}}\ and\ \bibinfo {author} {\bibfnamefont {I.}~\bibnamefont
  {Stewart}},\ }\href@noop {} {\bibfield  {journal} {\bibinfo  {journal}
  {Scientific american}\ }\textbf {\bibinfo {volume} {269}},\ \bibinfo {pages}
  {102} (\bibinfo {year} {1993})}\BibitemShut {NoStop}%
\bibitem [{\citenamefont {Pikovsky}\ and\ \citenamefont
  {Rosenblum}(2015)}]{pikovsky2015}%
  \BibitemOpen
  \bibfield  {author} {\bibinfo {author} {\bibfnamefont {A.}~\bibnamefont
  {Pikovsky}}\ and\ \bibinfo {author} {\bibfnamefont {M.}~\bibnamefont
  {Rosenblum}},\ }\href@noop {} {\bibfield  {journal} {\bibinfo  {journal}
  {Chaos: An Interdisciplinary Journal of Nonlinear Science}\ }\textbf
  {\bibinfo {volume} {25}},\ \bibinfo {pages} {097616} (\bibinfo {year}
  {2015})}\BibitemShut {NoStop}%
\bibitem [{\citenamefont {Abbott}\ and\ \citenamefont
  {Van~Vreeswijk}(1993)}]{abbott1993}%
  \BibitemOpen
  \bibfield  {author} {\bibinfo {author} {\bibfnamefont {L.~F.}\ \bibnamefont
  {Abbott}}\ and\ \bibinfo {author} {\bibfnamefont {C.}~\bibnamefont
  {Van~Vreeswijk}},\ }\href@noop {} {\bibfield  {journal} {\bibinfo  {journal}
  {Physical Review E}\ }\textbf {\bibinfo {volume} {48}},\ \bibinfo {pages}
  {1483} (\bibinfo {year} {1993})}\BibitemShut {NoStop}%
\bibitem [{\citenamefont {van Vreeswijk}(1996)}]{van1996}%
  \BibitemOpen
  \bibfield  {author} {\bibinfo {author} {\bibfnamefont {C.}~\bibnamefont {van
  Vreeswijk}},\ }\href@noop {} {\bibfield  {journal} {\bibinfo  {journal}
  {Physical Review E}\ }\textbf {\bibinfo {volume} {54}},\ \bibinfo {pages}
  {5522} (\bibinfo {year} {1996})}\BibitemShut {NoStop}%
\bibitem [{\citenamefont {Timme}\ \emph
  {et~al.}(2002{\natexlab{a}})\citenamefont {Timme}, \citenamefont {Wolf},\
  and\ \citenamefont {Geisel}}]{timme2002}%
  \BibitemOpen
  \bibfield  {author} {\bibinfo {author} {\bibfnamefont {M.}~\bibnamefont
  {Timme}}, \bibinfo {author} {\bibfnamefont {F.}~\bibnamefont {Wolf}}, \ and\
  \bibinfo {author} {\bibfnamefont {T.}~\bibnamefont {Geisel}},\ }\href@noop {}
  {\bibfield  {journal} {\bibinfo  {journal} {Physical review letters}\
  }\textbf {\bibinfo {volume} {89}},\ \bibinfo {pages} {258701} (\bibinfo
  {year} {2002}{\natexlab{a}})}\BibitemShut {NoStop}%
\bibitem [{\citenamefont {Vogels}\ \emph {et~al.}(2005)\citenamefont {Vogels},
  \citenamefont {Rajan},\ and\ \citenamefont {Abbott}}]{vogels2005}%
  \BibitemOpen
  \bibfield  {author} {\bibinfo {author} {\bibfnamefont {T.~P.}\ \bibnamefont
  {Vogels}}, \bibinfo {author} {\bibfnamefont {K.}~\bibnamefont {Rajan}}, \
  and\ \bibinfo {author} {\bibfnamefont {L.~F.}\ \bibnamefont {Abbott}},\
  }\href@noop {} {\bibfield  {journal} {\bibinfo  {journal} {Annu. Rev.
  Neurosci.}\ }\textbf {\bibinfo {volume} {28}},\ \bibinfo {pages} {357}
  (\bibinfo {year} {2005})}\BibitemShut {NoStop}%
\bibitem [{\citenamefont {Haken}(2006)}]{haken2006}%
  \BibitemOpen
  \bibfield  {author} {\bibinfo {author} {\bibfnamefont {H.}~\bibnamefont
  {Haken}},\ }\href@noop {} {\emph {\bibinfo {title} {Brain dynamics:
  synchronization and activity patterns in pulse-coupled neural nets with
  delays and noise}}}\ (\bibinfo  {publisher} {Springer Science \& Business
  Media},\ \bibinfo {year} {2006})\BibitemShut {NoStop}%
\bibitem [{\citenamefont {Olmi}\ \emph {et~al.}(2011)\citenamefont {Olmi},
  \citenamefont {Politi},\ and\ \citenamefont {Torcini}}]{olmi2011}%
  \BibitemOpen
  \bibfield  {author} {\bibinfo {author} {\bibfnamefont {S.}~\bibnamefont
  {Olmi}}, \bibinfo {author} {\bibfnamefont {A.}~\bibnamefont {Politi}}, \ and\
  \bibinfo {author} {\bibfnamefont {A.}~\bibnamefont {Torcini}},\ }\href@noop
  {} {\bibfield  {journal} {\bibinfo  {journal} {Europhysics Letters}\ }\textbf
  {\bibinfo {volume} {92}},\ \bibinfo {pages} {60007} (\bibinfo {year}
  {2011})}\BibitemShut {NoStop}%
\bibitem [{\citenamefont {Wang}\ and\ \citenamefont
  {Buzs{\'a}ki}(1996)}]{wang1996}%
  \BibitemOpen
  \bibfield  {author} {\bibinfo {author} {\bibfnamefont {X.-J.}\ \bibnamefont
  {Wang}}\ and\ \bibinfo {author} {\bibfnamefont {G.}~\bibnamefont
  {Buzs{\'a}ki}},\ }\href@noop {} {\bibfield  {journal} {\bibinfo  {journal}
  {Journal of neuroscience}\ }\textbf {\bibinfo {volume} {16}},\ \bibinfo
  {pages} {6402} (\bibinfo {year} {1996})}\BibitemShut {NoStop}%
\bibitem [{\citenamefont {Luccioli}\ and\ \citenamefont
  {Politi}(2010)}]{luccioli2010}%
  \BibitemOpen
  \bibfield  {author} {\bibinfo {author} {\bibfnamefont {S.}~\bibnamefont
  {Luccioli}}\ and\ \bibinfo {author} {\bibfnamefont {A.}~\bibnamefont
  {Politi}},\ }\href {\doibase 10.1103/PhysRevLett.105.158104} {\bibfield
  {journal} {\bibinfo  {journal} {Phys. Rev. Lett.}\ }\textbf {\bibinfo
  {volume} {105}},\ \bibinfo {pages} {158104} (\bibinfo {year}
  {2010})}\BibitemShut {NoStop}%
\bibitem [{\citenamefont {Ullner}\ and\ \citenamefont
  {Politi}(2016)}]{ullner2016}%
  \BibitemOpen
  \bibfield  {author} {\bibinfo {author} {\bibfnamefont {E.}~\bibnamefont
  {Ullner}}\ and\ \bibinfo {author} {\bibfnamefont {A.}~\bibnamefont
  {Politi}},\ }\href@noop {} {\bibfield  {journal} {\bibinfo  {journal}
  {Physical Review X}\ }\textbf {\bibinfo {volume} {6}},\ \bibinfo {pages}
  {011015} (\bibinfo {year} {2016})}\BibitemShut {NoStop}%
\bibitem [{\citenamefont {Softky}\ and\ \citenamefont
  {Koch}(1993)}]{softky1993}%
  \BibitemOpen
  \bibfield  {author} {\bibinfo {author} {\bibfnamefont {W.~R.}\ \bibnamefont
  {Softky}}\ and\ \bibinfo {author} {\bibfnamefont {C.}~\bibnamefont {Koch}},\
  }\href@noop {} {\bibfield  {journal} {\bibinfo  {journal} {Journal of
  neuroscience}\ }\textbf {\bibinfo {volume} {13}},\ \bibinfo {pages} {334}
  (\bibinfo {year} {1993})}\BibitemShut {NoStop}%
\bibitem [{\citenamefont {Destexhe}\ and\ \citenamefont
  {Par{\'e}}(1999)}]{destexhe1999}%
  \BibitemOpen
  \bibfield  {author} {\bibinfo {author} {\bibfnamefont {A.}~\bibnamefont
  {Destexhe}}\ and\ \bibinfo {author} {\bibfnamefont {D.}~\bibnamefont
  {Par{\'e}}},\ }\href@noop {} {\bibfield  {journal} {\bibinfo  {journal}
  {Journal of neurophysiology}\ }\textbf {\bibinfo {volume} {81}},\ \bibinfo
  {pages} {1531} (\bibinfo {year} {1999})}\BibitemShut {NoStop}%
\bibitem [{\citenamefont {Bruno}\ and\ \citenamefont
  {Sakmann}(2006)}]{bruno2006}%
  \BibitemOpen
  \bibfield  {author} {\bibinfo {author} {\bibfnamefont {R.~M.}\ \bibnamefont
  {Bruno}}\ and\ \bibinfo {author} {\bibfnamefont {B.}~\bibnamefont
  {Sakmann}},\ }\href@noop {} {\bibfield  {journal} {\bibinfo  {journal}
  {Science}\ }\textbf {\bibinfo {volume} {312}},\ \bibinfo {pages} {1622}
  (\bibinfo {year} {2006})}\BibitemShut {NoStop}%
\bibitem [{\citenamefont {Lefort}\ \emph {et~al.}(2009)\citenamefont {Lefort},
  \citenamefont {Tomm}, \citenamefont {Sarria},\ and\ \citenamefont
  {Petersen}}]{lefort2009}%
  \BibitemOpen
  \bibfield  {author} {\bibinfo {author} {\bibfnamefont {S.}~\bibnamefont
  {Lefort}}, \bibinfo {author} {\bibfnamefont {C.}~\bibnamefont {Tomm}},
  \bibinfo {author} {\bibfnamefont {J.-C.~F.}\ \bibnamefont {Sarria}}, \ and\
  \bibinfo {author} {\bibfnamefont {C.~C.}\ \bibnamefont {Petersen}},\ }\href
  {\doibase https://doi.org/10.1016/j.neuron.2008.12.020} {\bibfield  {journal}
  {\bibinfo  {journal} {Neuron}\ }\textbf {\bibinfo {volume} {61}},\ \bibinfo
  {pages} {301 } (\bibinfo {year} {2009})}\BibitemShut {NoStop}%
\bibitem [{\citenamefont {Shadlen}\ and\ \citenamefont
  {Newsome}(1994)}]{shadlen1994}%
  \BibitemOpen
  \bibfield  {author} {\bibinfo {author} {\bibfnamefont {M.~N.}\ \bibnamefont
  {Shadlen}}\ and\ \bibinfo {author} {\bibfnamefont {W.~T.}\ \bibnamefont
  {Newsome}},\ }\href@noop {} {\bibfield  {journal} {\bibinfo  {journal}
  {Current opinion in neurobiology}\ }\textbf {\bibinfo {volume} {4}},\
  \bibinfo {pages} {569} (\bibinfo {year} {1994})}\BibitemShut {NoStop}%
\bibitem [{\citenamefont {Brunel}\ and\ \citenamefont
  {Hakim}(1999)}]{brunel1999}%
  \BibitemOpen
  \bibfield  {author} {\bibinfo {author} {\bibfnamefont {N.}~\bibnamefont
  {Brunel}}\ and\ \bibinfo {author} {\bibfnamefont {V.}~\bibnamefont {Hakim}},\
  }\href@noop {} {\bibfield  {journal} {\bibinfo  {journal} {Neural
  computation}\ }\textbf {\bibinfo {volume} {11}},\ \bibinfo {pages} {1621}
  (\bibinfo {year} {1999})}\BibitemShut {NoStop}%
\bibitem [{\citenamefont {Barral}\ and\ \citenamefont
  {Reyes}(2016)}]{barral2016}%
  \BibitemOpen
  \bibfield  {author} {\bibinfo {author} {\bibfnamefont {J.}~\bibnamefont
  {Barral}}\ and\ \bibinfo {author} {\bibfnamefont {A.~D.}\ \bibnamefont
  {Reyes}},\ }\href@noop {} {\bibfield  {journal} {\bibinfo  {journal} {Nature
  neuroscience}\ }\textbf {\bibinfo {volume} {19}},\ \bibinfo {pages} {1690}
  (\bibinfo {year} {2016})}\BibitemShut {NoStop}%
\bibitem [{\citenamefont {Renart}\ \emph {et~al.}(2010)\citenamefont {Renart},
  \citenamefont {de~la Rocha}, \citenamefont {Bartho}, \citenamefont
  {Hollender}, \citenamefont {Parga}, \citenamefont {Reyes},\ and\
  \citenamefont {Harris}}]{bal2}%
  \BibitemOpen
  \bibfield  {author} {\bibinfo {author} {\bibfnamefont {A.}~\bibnamefont
  {Renart}}, \bibinfo {author} {\bibfnamefont {J.}~\bibnamefont {de~la Rocha}},
  \bibinfo {author} {\bibfnamefont {P.}~\bibnamefont {Bartho}}, \bibinfo
  {author} {\bibfnamefont {L.}~\bibnamefont {Hollender}}, \bibinfo {author}
  {\bibfnamefont {N.}~\bibnamefont {Parga}}, \bibinfo {author} {\bibfnamefont
  {A.}~\bibnamefont {Reyes}}, \ and\ \bibinfo {author} {\bibfnamefont {K.~D.}\
  \bibnamefont {Harris}},\ }\href {\doibase 10.1126/science.1179850} {\bibfield
   {journal} {\bibinfo  {journal} {Science}\ }\textbf {\bibinfo {volume}
  {327}},\ \bibinfo {pages} {587} (\bibinfo {year} {2010})}\BibitemShut
  {NoStop}%
\bibitem [{\citenamefont {Monteforte}\ and\ \citenamefont {Wolf}(2010)}]{wolf}%
  \BibitemOpen
  \bibfield  {author} {\bibinfo {author} {\bibfnamefont {M.}~\bibnamefont
  {Monteforte}}\ and\ \bibinfo {author} {\bibfnamefont {F.}~\bibnamefont
  {Wolf}},\ }\href {\doibase 10.1103/PhysRevLett.105.268104} {\bibfield
  {journal} {\bibinfo  {journal} {Phys. Rev. Lett.}\ }\textbf {\bibinfo
  {volume} {105}},\ \bibinfo {pages} {268104} (\bibinfo {year}
  {2010})}\BibitemShut {NoStop}%
\bibitem [{\citenamefont {Litwin-Kumar}\ and\ \citenamefont
  {Doiron}(2012)}]{bal3}%
  \BibitemOpen
  \bibfield  {author} {\bibinfo {author} {\bibfnamefont {A.}~\bibnamefont
  {Litwin-Kumar}}\ and\ \bibinfo {author} {\bibfnamefont {B.}~\bibnamefont
  {Doiron}},\ }\href {http://dx.doi.org/10.1038/nn.3220} {\bibfield  {journal}
  {\bibinfo  {journal} {Nat Neurosci}\ }\textbf {\bibinfo {volume} {15}},\
  \bibinfo {pages} {1498} (\bibinfo {year} {2012})}\BibitemShut {NoStop}%
\bibitem [{\citenamefont {Kadmon}\ and\ \citenamefont
  {Sompolinsky}(2015)}]{bal4}%
  \BibitemOpen
  \bibfield  {author} {\bibinfo {author} {\bibfnamefont {J.}~\bibnamefont
  {Kadmon}}\ and\ \bibinfo {author} {\bibfnamefont {H.}~\bibnamefont
  {Sompolinsky}},\ }\href {\doibase 10.1103/PhysRevX.5.041030} {\bibfield
  {journal} {\bibinfo  {journal} {Phys. Rev. X}\ }\textbf {\bibinfo {volume}
  {5}},\ \bibinfo {pages} {041030} (\bibinfo {year} {2015})}\BibitemShut
  {NoStop}%
\bibitem [{\citenamefont {Rosenbaum}\ and\ \citenamefont
  {Doiron}(2014)}]{rosenbaum2014}%
  \BibitemOpen
  \bibfield  {author} {\bibinfo {author} {\bibfnamefont {R.}~\bibnamefont
  {Rosenbaum}}\ and\ \bibinfo {author} {\bibfnamefont {B.}~\bibnamefont
  {Doiron}},\ }\href@noop {} {\bibfield  {journal} {\bibinfo  {journal}
  {Physical Review X}\ }\textbf {\bibinfo {volume} {4}},\ \bibinfo {pages}
  {021039} (\bibinfo {year} {2014})}\BibitemShut {NoStop}%
\bibitem [{\citenamefont {Pyle}\ and\ \citenamefont
  {Rosenbaum}(2016)}]{pyle2016}%
  \BibitemOpen
  \bibfield  {author} {\bibinfo {author} {\bibfnamefont {R.}~\bibnamefont
  {Pyle}}\ and\ \bibinfo {author} {\bibfnamefont {R.}~\bibnamefont
  {Rosenbaum}},\ }\href@noop {} {\bibfield  {journal} {\bibinfo  {journal}
  {Physical Review E}\ }\textbf {\bibinfo {volume} {93}},\ \bibinfo {pages}
  {040302} (\bibinfo {year} {2016})}\BibitemShut {NoStop}%
\bibitem [{\citenamefont {di~Volo}\ and\ \citenamefont
  {Torcini}(2018)}]{matteo}%
  \BibitemOpen
  \bibfield  {author} {\bibinfo {author} {\bibfnamefont {M.}~\bibnamefont
  {di~Volo}}\ and\ \bibinfo {author} {\bibfnamefont {A.}~\bibnamefont
  {Torcini}},\ }\href@noop {} {\bibfield  {journal} {\bibinfo  {journal} {Phys.
  Rev. Lett.}\ }\textbf {\bibinfo {volume} {121}},\ \bibinfo {pages} {128301}
  (\bibinfo {year} {2018})}\BibitemShut {NoStop}%
\bibitem [{\citenamefont {Di~Volo}\ \emph {et~al.}(2022)\citenamefont
  {Di~Volo}, \citenamefont {Segneri}, \citenamefont {Goldobin}, \citenamefont
  {Politi},\ and\ \citenamefont {Torcini}}]{noi}%
  \BibitemOpen
  \bibfield  {author} {\bibinfo {author} {\bibfnamefont {M.}~\bibnamefont
  {Di~Volo}}, \bibinfo {author} {\bibfnamefont {M.}~\bibnamefont {Segneri}},
  \bibinfo {author} {\bibfnamefont {D.~S.}\ \bibnamefont {Goldobin}}, \bibinfo
  {author} {\bibfnamefont {A.}~\bibnamefont {Politi}}, \ and\ \bibinfo {author}
  {\bibfnamefont {A.}~\bibnamefont {Torcini}},\ }\href@noop {} {\bibfield
  {journal} {\bibinfo  {journal} {Chaos: An Interdisciplinary Journal of
  Nonlinear Science}\ }\textbf {\bibinfo {volume} {32}},\ \bibinfo {pages}
  {023120} (\bibinfo {year} {2022})}\BibitemShut {NoStop}%
\bibitem [{\citenamefont {Ahmadian}\ and\ \citenamefont
  {Miller}(2021)}]{ahmadian2021}%
  \BibitemOpen
  \bibfield  {author} {\bibinfo {author} {\bibfnamefont {Y.}~\bibnamefont
  {Ahmadian}}\ and\ \bibinfo {author} {\bibfnamefont {K.~D.}\ \bibnamefont
  {Miller}},\ }\href@noop {} {\bibfield  {journal} {\bibinfo  {journal}
  {Neuron}\ }\textbf {\bibinfo {volume} {109}},\ \bibinfo {pages} {3373}
  (\bibinfo {year} {2021})}\BibitemShut {NoStop}%
\bibitem [{\citenamefont {Khajeh}\ \emph {et~al.}(2022)\citenamefont {Khajeh},
  \citenamefont {Fumarola},\ and\ \citenamefont {Abbott}}]{khajeh2022}%
  \BibitemOpen
  \bibfield  {author} {\bibinfo {author} {\bibfnamefont {R.}~\bibnamefont
  {Khajeh}}, \bibinfo {author} {\bibfnamefont {F.}~\bibnamefont {Fumarola}}, \
  and\ \bibinfo {author} {\bibfnamefont {L.}~\bibnamefont {Abbott}},\
  }\href@noop {} {\bibfield  {journal} {\bibinfo  {journal} {PLOS Computational
  Biology}\ }\textbf {\bibinfo {volume} {18}},\ \bibinfo {pages} {e1008836}
  (\bibinfo {year} {2022})}\BibitemShut {NoStop}%
\bibitem [{\citenamefont {Chung}\ and\ \citenamefont
  {Ferster}(1998)}]{chung1998}%
  \BibitemOpen
  \bibfield  {author} {\bibinfo {author} {\bibfnamefont {S.}~\bibnamefont
  {Chung}}\ and\ \bibinfo {author} {\bibfnamefont {D.}~\bibnamefont
  {Ferster}},\ }\href@noop {} {\bibfield  {journal} {\bibinfo  {journal}
  {Neuron}\ }\textbf {\bibinfo {volume} {20}},\ \bibinfo {pages} {1177}
  (\bibinfo {year} {1998})}\BibitemShut {NoStop}%
\bibitem [{\citenamefont {Finn}\ \emph {et~al.}(2007)\citenamefont {Finn},
  \citenamefont {Priebe},\ and\ \citenamefont {Ferster}}]{finn2007}%
  \BibitemOpen
  \bibfield  {author} {\bibinfo {author} {\bibfnamefont {I.~M.}\ \bibnamefont
  {Finn}}, \bibinfo {author} {\bibfnamefont {N.~J.}\ \bibnamefont {Priebe}}, \
  and\ \bibinfo {author} {\bibfnamefont {D.}~\bibnamefont {Ferster}},\
  }\href@noop {} {\bibfield  {journal} {\bibinfo  {journal} {Neuron}\ }\textbf
  {\bibinfo {volume} {54}},\ \bibinfo {pages} {137} (\bibinfo {year}
  {2007})}\BibitemShut {NoStop}%
\bibitem [{\citenamefont {Lien}\ and\ \citenamefont
  {Scanziani}(2013)}]{lien2013}%
  \BibitemOpen
  \bibfield  {author} {\bibinfo {author} {\bibfnamefont {A.~D.}\ \bibnamefont
  {Lien}}\ and\ \bibinfo {author} {\bibfnamefont {M.}~\bibnamefont
  {Scanziani}},\ }\href@noop {} {\bibfield  {journal} {\bibinfo  {journal}
  {Nature neuroscience}\ }\textbf {\bibinfo {volume} {16}},\ \bibinfo {pages}
  {1315} (\bibinfo {year} {2013})}\BibitemShut {NoStop}%
\bibitem [{\citenamefont {Tsodyks}\ and\ \citenamefont
  {Wu}(2013)}]{tsodyks2013}%
  \BibitemOpen
  \bibfield  {author} {\bibinfo {author} {\bibfnamefont {M.}~\bibnamefont
  {Tsodyks}}\ and\ \bibinfo {author} {\bibfnamefont {S.}~\bibnamefont {Wu}},\
  }\href@noop {} {\bibfield  {journal} {\bibinfo  {journal} {Scholarpedia}\
  }\textbf {\bibinfo {volume} {8}},\ \bibinfo {pages} {3153} (\bibinfo {year}
  {2013})}\BibitemShut {NoStop}%
\bibitem [{\citenamefont {Tsodyks}\ and\ \citenamefont
  {Markram}(1997)}]{tsodyks1997}%
  \BibitemOpen
  \bibfield  {author} {\bibinfo {author} {\bibfnamefont {M.~V.}\ \bibnamefont
  {Tsodyks}}\ and\ \bibinfo {author} {\bibfnamefont {H.}~\bibnamefont
  {Markram}},\ }\href@noop {} {\bibfield  {journal} {\bibinfo  {journal}
  {Proceedings of the national academy of sciences}\ }\textbf {\bibinfo
  {volume} {94}},\ \bibinfo {pages} {719} (\bibinfo {year} {1997})}\BibitemShut
  {NoStop}%
\bibitem [{\citenamefont {Varela}\ \emph {et~al.}(1999)\citenamefont {Varela},
  \citenamefont {Song}, \citenamefont {Turrigiano},\ and\ \citenamefont
  {Nelson}}]{varela1999}%
  \BibitemOpen
  \bibfield  {author} {\bibinfo {author} {\bibfnamefont {J.~A.}\ \bibnamefont
  {Varela}}, \bibinfo {author} {\bibfnamefont {S.}~\bibnamefont {Song}},
  \bibinfo {author} {\bibfnamefont {G.~G.}\ \bibnamefont {Turrigiano}}, \ and\
  \bibinfo {author} {\bibfnamefont {S.~B.}\ \bibnamefont {Nelson}},\
  }\href@noop {} {\bibfield  {journal} {\bibinfo  {journal} {Journal of
  Neuroscience}\ }\textbf {\bibinfo {volume} {19}},\ \bibinfo {pages} {4293}
  (\bibinfo {year} {1999})}\BibitemShut {NoStop}%
\bibitem [{\citenamefont {Varela}\ \emph {et~al.}(1997)\citenamefont {Varela},
  \citenamefont {Sen}, \citenamefont {Gibson}, \citenamefont {Fost},
  \citenamefont {Abbott},\ and\ \citenamefont {Nelson}}]{varela1997}%
  \BibitemOpen
  \bibfield  {author} {\bibinfo {author} {\bibfnamefont {J.~A.}\ \bibnamefont
  {Varela}}, \bibinfo {author} {\bibfnamefont {K.}~\bibnamefont {Sen}},
  \bibinfo {author} {\bibfnamefont {J.}~\bibnamefont {Gibson}}, \bibinfo
  {author} {\bibfnamefont {J.}~\bibnamefont {Fost}}, \bibinfo {author}
  {\bibfnamefont {L.}~\bibnamefont {Abbott}}, \ and\ \bibinfo {author}
  {\bibfnamefont {S.~B.}\ \bibnamefont {Nelson}},\ }\href@noop {} {\bibfield
  {journal} {\bibinfo  {journal} {Journal of Neuroscience}\ }\textbf {\bibinfo
  {volume} {17}},\ \bibinfo {pages} {7926} (\bibinfo {year}
  {1997})}\BibitemShut {NoStop}%
\bibitem [{\citenamefont {Van~Vreeswijk}\ \emph {et~al.}(1994)\citenamefont
  {Van~Vreeswijk}, \citenamefont {Abbott},\ and\ \citenamefont
  {Ermentrout}}]{van1994}%
  \BibitemOpen
  \bibfield  {author} {\bibinfo {author} {\bibfnamefont {C.}~\bibnamefont
  {Van~Vreeswijk}}, \bibinfo {author} {\bibfnamefont {L.}~\bibnamefont
  {Abbott}}, \ and\ \bibinfo {author} {\bibfnamefont {G.~B.}\ \bibnamefont
  {Ermentrout}},\ }\href@noop {} {\bibfield  {journal} {\bibinfo  {journal}
  {Journal of computational neuroscience}\ }\textbf {\bibinfo {volume} {1}},\
  \bibinfo {pages} {313} (\bibinfo {year} {1994})}\BibitemShut {NoStop}%
\bibitem [{\citenamefont {Coombes}\ \emph {et~al.}(2003)\citenamefont
  {Coombes}, \citenamefont {Lord},\ and\ \citenamefont {Owen}}]{coombes2003}%
  \BibitemOpen
  \bibfield  {author} {\bibinfo {author} {\bibfnamefont {S.}~\bibnamefont
  {Coombes}}, \bibinfo {author} {\bibfnamefont {G.~J.}\ \bibnamefont {Lord}}, \
  and\ \bibinfo {author} {\bibfnamefont {M.~R.}\ \bibnamefont {Owen}},\
  }\href@noop {} {\bibfield  {journal} {\bibinfo  {journal} {Physica D:
  Nonlinear Phenomena}\ }\textbf {\bibinfo {volume} {178}},\ \bibinfo {pages}
  {219} (\bibinfo {year} {2003})}\BibitemShut {NoStop}%
\bibitem [{\citenamefont {Mirollo}\ and\ \citenamefont
  {Strogatz}(1990)}]{mirollo1990}%
  \BibitemOpen
  \bibfield  {author} {\bibinfo {author} {\bibfnamefont {R.~E.}\ \bibnamefont
  {Mirollo}}\ and\ \bibinfo {author} {\bibfnamefont {S.~H.}\ \bibnamefont
  {Strogatz}},\ }\href@noop {} {\bibfield  {journal} {\bibinfo  {journal} {SIAM
  Journal on Applied Mathematics}\ }\textbf {\bibinfo {volume} {50}},\ \bibinfo
  {pages} {1645} (\bibinfo {year} {1990})}\BibitemShut {NoStop}%
\bibitem [{\citenamefont {Timme}\ \emph
  {et~al.}(2002{\natexlab{b}})\citenamefont {Timme}, \citenamefont {Wolf},\
  and\ \citenamefont {Geisel}}]{timme}%
  \BibitemOpen
  \bibfield  {author} {\bibinfo {author} {\bibfnamefont {M.}~\bibnamefont
  {Timme}}, \bibinfo {author} {\bibfnamefont {F.}~\bibnamefont {Wolf}}, \ and\
  \bibinfo {author} {\bibfnamefont {T.}~\bibnamefont {Geisel}},\ }\href
  {\doibase 10.1103/PhysRevLett.89.258701} {\bibfield  {journal} {\bibinfo
  {journal} {Phys. Rev. Lett.}\ }\textbf {\bibinfo {volume} {89}},\ \bibinfo
  {pages} {258701} (\bibinfo {year} {2002}{\natexlab{b}})}\BibitemShut
  {NoStop}%
\bibitem [{\citenamefont {Burkitt}(2006)}]{burkitt2006}%
  \BibitemOpen
  \bibfield  {author} {\bibinfo {author} {\bibfnamefont {A.~N.}\ \bibnamefont
  {Burkitt}},\ }\href@noop {} {\bibfield  {journal} {\bibinfo  {journal}
  {Biological cybernetics}\ }\textbf {\bibinfo {volume} {95}},\ \bibinfo
  {pages} {1} (\bibinfo {year} {2006})}\BibitemShut {NoStop}%
\bibitem [{\citenamefont {Smeal}\ \emph {et~al.}(2010)\citenamefont {Smeal},
  \citenamefont {Ermentrout},\ and\ \citenamefont {White}}]{smeal2010}%
  \BibitemOpen
  \bibfield  {author} {\bibinfo {author} {\bibfnamefont {R.~M.}\ \bibnamefont
  {Smeal}}, \bibinfo {author} {\bibfnamefont {G.~B.}\ \bibnamefont
  {Ermentrout}}, \ and\ \bibinfo {author} {\bibfnamefont {J.~A.}\ \bibnamefont
  {White}},\ }\href@noop {} {\bibfield  {journal} {\bibinfo  {journal}
  {Philosophical Transactions of the Royal Society B: Biological Sciences}\
  }\textbf {\bibinfo {volume} {365}},\ \bibinfo {pages} {2407} (\bibinfo {year}
  {2010})}\BibitemShut {NoStop}%
\bibitem [{\citenamefont {Ermentrout}(1996)}]{ermentrout1996}%
  \BibitemOpen
  \bibfield  {author} {\bibinfo {author} {\bibfnamefont {B.}~\bibnamefont
  {Ermentrout}},\ }\href@noop {} {\bibfield  {journal} {\bibinfo  {journal}
  {Neural computation}\ }\textbf {\bibinfo {volume} {8}},\ \bibinfo {pages}
  {979} (\bibinfo {year} {1996})}\BibitemShut {NoStop}%
\bibitem [{\citenamefont {Tsodyks}\ \emph {et~al.}(1998)\citenamefont
  {Tsodyks}, \citenamefont {Pawelzik},\ and\ \citenamefont
  {Markram}}]{tsodyks1998}%
  \BibitemOpen
  \bibfield  {author} {\bibinfo {author} {\bibfnamefont {M.}~\bibnamefont
  {Tsodyks}}, \bibinfo {author} {\bibfnamefont {K.}~\bibnamefont {Pawelzik}}, \
  and\ \bibinfo {author} {\bibfnamefont {H.}~\bibnamefont {Markram}},\
  }\href@noop {} {\bibfield  {journal} {\bibinfo  {journal} {Neural
  computation}\ }\textbf {\bibinfo {volume} {10}},\ \bibinfo {pages} {821}
  (\bibinfo {year} {1998})}\BibitemShut {NoStop}%
\bibitem [{\citenamefont {Golomb}\ \emph {et~al.}(2001)\citenamefont {Golomb},
  \citenamefont {Hansel},\ and\ \citenamefont {Mato}}]{golomb2001}%
  \BibitemOpen
  \bibfield  {author} {\bibinfo {author} {\bibfnamefont {D.}~\bibnamefont
  {Golomb}}, \bibinfo {author} {\bibfnamefont {D.}~\bibnamefont {Hansel}}, \
  and\ \bibinfo {author} {\bibfnamefont {G.}~\bibnamefont {Mato}},\ }\href@noop
  {} {\bibfield  {journal} {\bibinfo  {journal} {in Handbook of Biological
  Physics}\ }\textbf {\bibinfo {volume} {4}},\ \bibinfo {pages} {887} (\bibinfo
  {year} {2001})}\BibitemShut {NoStop}%
\bibitem [{par()}]{param}%
  \BibitemOpen
  \href@noop {} {}\bibinfo {note} {The parameters are set as follows. Each
  neuron has a probability of 10\% to be connected to any other neuron, to
  guarantee that 80\% (20\%) of these connections are excitatory (inhibitory)
  as in the cortex we set $p^e = 0.08$ ($p^i = 0.02$). For the STD parameters
  we set $u =0.5$ (a single spike emission halves the synaptic resources) and
  $\tau_d = 1$ s. The overall coupling strength has been fixed to $G=1$. For
  the $Z_{\rm I}(\phi)$ ($Z_{\rm LIF}(\phi)$) PRC we considered synapses with
  $\alpha^{-1} = 0.2$ ms ($\alpha^{-1} = 0.04$ ms).}\BibitemShut {Stop}%
\bibitem [{not()}]{note1}%
  \BibitemOpen
  \href@noop {} {}\bibinfo {note} {In our case, the inhibitory field is equal
  to the excitatory field.}\BibitemShut {Stop}%
\bibitem [{\citenamefont {Hrom{\'a}dka}\ \emph {et~al.}(2008)\citenamefont
  {Hrom{\'a}dka}, \citenamefont {DeWeese},\ and\ \citenamefont
  {Zador}}]{hromadka2008}%
  \BibitemOpen
  \bibfield  {author} {\bibinfo {author} {\bibfnamefont {T.}~\bibnamefont
  {Hrom{\'a}dka}}, \bibinfo {author} {\bibfnamefont {M.~R.}\ \bibnamefont
  {DeWeese}}, \ and\ \bibinfo {author} {\bibfnamefont {A.~M.}\ \bibnamefont
  {Zador}},\ }\href@noop {} {\bibfield  {journal} {\bibinfo  {journal} {PLoS
  biology}\ }\textbf {\bibinfo {volume} {6}},\ \bibinfo {pages} {e16} (\bibinfo
  {year} {2008})}\BibitemShut {NoStop}%
\bibitem [{\citenamefont {O'Connor}\ \emph {et~al.}(2010)\citenamefont
  {O'Connor}, \citenamefont {Peron}, \citenamefont {Huber},\ and\ \citenamefont
  {Svoboda}}]{o2010}%
  \BibitemOpen
  \bibfield  {author} {\bibinfo {author} {\bibfnamefont {D.~H.}\ \bibnamefont
  {O'Connor}}, \bibinfo {author} {\bibfnamefont {S.~P.}\ \bibnamefont {Peron}},
  \bibinfo {author} {\bibfnamefont {D.}~\bibnamefont {Huber}}, \ and\ \bibinfo
  {author} {\bibfnamefont {K.}~\bibnamefont {Svoboda}},\ }\href@noop {}
  {\bibfield  {journal} {\bibinfo  {journal} {Neuron}\ }\textbf {\bibinfo
  {volume} {67}},\ \bibinfo {pages} {1048} (\bibinfo {year}
  {2010})}\BibitemShut {NoStop}%
\bibitem [{\citenamefont {Wohrer}\ \emph {et~al.}(2013)\citenamefont {Wohrer},
  \citenamefont {Humphries},\ and\ \citenamefont {Machens}}]{wohrer2013}%
  \BibitemOpen
  \bibfield  {author} {\bibinfo {author} {\bibfnamefont {A.}~\bibnamefont
  {Wohrer}}, \bibinfo {author} {\bibfnamefont {M.~D.}\ \bibnamefont
  {Humphries}}, \ and\ \bibinfo {author} {\bibfnamefont {C.~K.}\ \bibnamefont
  {Machens}},\ }\href@noop {} {\bibfield  {journal} {\bibinfo  {journal}
  {Progress in neurobiology}\ }\textbf {\bibinfo {volume} {103}},\ \bibinfo
  {pages} {156} (\bibinfo {year} {2013})}\BibitemShut {NoStop}%
\bibitem [{\citenamefont {Buzs{\'a}ki}\ and\ \citenamefont
  {Mizuseki}(2014)}]{buzsaki2014}%
  \BibitemOpen
  \bibfield  {author} {\bibinfo {author} {\bibfnamefont {G.}~\bibnamefont
  {Buzs{\'a}ki}}\ and\ \bibinfo {author} {\bibfnamefont {K.}~\bibnamefont
  {Mizuseki}},\ }\href@noop {} {\bibfield  {journal} {\bibinfo  {journal}
  {Nature Reviews Neuroscience}\ }\textbf {\bibinfo {volume} {15}},\ \bibinfo
  {pages} {264} (\bibinfo {year} {2014})}\BibitemShut {NoStop}%
\bibitem [{\citenamefont {Mongillo}\ \emph {et~al.}(2018)\citenamefont
  {Mongillo}, \citenamefont {Rumpel},\ and\ \citenamefont
  {Loewenstein}}]{mongillo2018}%
  \BibitemOpen
  \bibfield  {author} {\bibinfo {author} {\bibfnamefont {G.}~\bibnamefont
  {Mongillo}}, \bibinfo {author} {\bibfnamefont {S.}~\bibnamefont {Rumpel}}, \
  and\ \bibinfo {author} {\bibfnamefont {Y.}~\bibnamefont {Loewenstein}},\
  }\href@noop {} {\bibfield  {journal} {\bibinfo  {journal} {Nature
  neuroscience}\ }\textbf {\bibinfo {volume} {21}},\ \bibinfo {pages} {1463}
  (\bibinfo {year} {2018})}\BibitemShut {NoStop}%
\bibitem [{\citenamefont {Roxin}\ \emph {et~al.}(2011)\citenamefont {Roxin},
  \citenamefont {Brunel}, \citenamefont {Hansel}, \citenamefont {Mongillo},\
  and\ \citenamefont {van Vreeswijk}}]{roxin2011}%
  \BibitemOpen
  \bibfield  {author} {\bibinfo {author} {\bibfnamefont {A.}~\bibnamefont
  {Roxin}}, \bibinfo {author} {\bibfnamefont {N.}~\bibnamefont {Brunel}},
  \bibinfo {author} {\bibfnamefont {D.}~\bibnamefont {Hansel}}, \bibinfo
  {author} {\bibfnamefont {G.}~\bibnamefont {Mongillo}}, \ and\ \bibinfo
  {author} {\bibfnamefont {C.}~\bibnamefont {van Vreeswijk}},\ }\href@noop {}
  {\bibfield  {journal} {\bibinfo  {journal} {Journal of Neuroscience}\
  }\textbf {\bibinfo {volume} {31}},\ \bibinfo {pages} {16217} (\bibinfo {year}
  {2011})}\BibitemShut {NoStop}%
\bibitem [{\citenamefont {Del~Giudice}\ \emph {et~al.}(2003)\citenamefont
  {Del~Giudice}, \citenamefont {Fusi},\ and\ \citenamefont {Mattia}}]{del2003}%
  \BibitemOpen
  \bibfield  {author} {\bibinfo {author} {\bibfnamefont {P.}~\bibnamefont
  {Del~Giudice}}, \bibinfo {author} {\bibfnamefont {S.}~\bibnamefont {Fusi}}, \
  and\ \bibinfo {author} {\bibfnamefont {M.}~\bibnamefont {Mattia}},\
  }\href@noop {} {\bibfield  {journal} {\bibinfo  {journal} {Journal of
  Physiology-Paris}\ }\textbf {\bibinfo {volume} {97}},\ \bibinfo {pages} {659}
  (\bibinfo {year} {2003})}\BibitemShut {NoStop}%
\bibitem [{\citenamefont {Mongillo}\ \emph {et~al.}(2008)\citenamefont
  {Mongillo}, \citenamefont {Barak},\ and\ \citenamefont
  {Tsodyks}}]{mongillo2008}%
  \BibitemOpen
  \bibfield  {author} {\bibinfo {author} {\bibfnamefont {G.}~\bibnamefont
  {Mongillo}}, \bibinfo {author} {\bibfnamefont {O.}~\bibnamefont {Barak}}, \
  and\ \bibinfo {author} {\bibfnamefont {M.}~\bibnamefont {Tsodyks}},\
  }\href@noop {} {\bibfield  {journal} {\bibinfo  {journal} {Science}\ }\textbf
  {\bibinfo {volume} {319}},\ \bibinfo {pages} {1543} (\bibinfo {year}
  {2008})}\BibitemShut {NoStop}%
\bibitem [{\citenamefont {Taher}\ \emph {et~al.}(2020)\citenamefont {Taher},
  \citenamefont {Torcini},\ and\ \citenamefont {Olmi}}]{taher2020}%
  \BibitemOpen
  \bibfield  {author} {\bibinfo {author} {\bibfnamefont {H.}~\bibnamefont
  {Taher}}, \bibinfo {author} {\bibfnamefont {A.}~\bibnamefont {Torcini}}, \
  and\ \bibinfo {author} {\bibfnamefont {S.}~\bibnamefont {Olmi}},\ }\href@noop
  {} {\bibfield  {journal} {\bibinfo  {journal} {PLOS Computational Biology}\
  }\textbf {\bibinfo {volume} {16}},\ \bibinfo {pages} {e1008533} (\bibinfo
  {year} {2020})}\BibitemShut {NoStop}%
\bibitem [{\citenamefont {Romani}\ and\ \citenamefont
  {Tsodyks}(2015)}]{romani2015}%
  \BibitemOpen
  \bibfield  {author} {\bibinfo {author} {\bibfnamefont {S.}~\bibnamefont
  {Romani}}\ and\ \bibinfo {author} {\bibfnamefont {M.}~\bibnamefont
  {Tsodyks}},\ }\href@noop {} {\bibfield  {journal} {\bibinfo  {journal}
  {Hippocampus}\ }\textbf {\bibinfo {volume} {25}},\ \bibinfo {pages} {94}
  (\bibinfo {year} {2015})}\BibitemShut {NoStop}%
\bibitem [{\citenamefont {Wang}\ \emph {et~al.}(2015)\citenamefont {Wang},
  \citenamefont {Romani}, \citenamefont {Lustig}, \citenamefont {Leonardo},\
  and\ \citenamefont {Pastalkova}}]{wang2015}%
  \BibitemOpen
  \bibfield  {author} {\bibinfo {author} {\bibfnamefont {Y.}~\bibnamefont
  {Wang}}, \bibinfo {author} {\bibfnamefont {S.}~\bibnamefont {Romani}},
  \bibinfo {author} {\bibfnamefont {B.}~\bibnamefont {Lustig}}, \bibinfo
  {author} {\bibfnamefont {A.}~\bibnamefont {Leonardo}}, \ and\ \bibinfo
  {author} {\bibfnamefont {E.}~\bibnamefont {Pastalkova}},\ }\href@noop {}
  {\bibfield  {journal} {\bibinfo  {journal} {Nature neuroscience}\ }\textbf
  {\bibinfo {volume} {18}},\ \bibinfo {pages} {282} (\bibinfo {year}
  {2015})}\BibitemShut {NoStop}%
\bibitem [{\citenamefont {Haimerl}\ \emph {et~al.}(2019)\citenamefont
  {Haimerl}, \citenamefont {Angulo-Garcia}, \citenamefont {Villette},
  \citenamefont {Reichinnek}, \citenamefont {Torcini}, \citenamefont
  {Cossart},\ and\ \citenamefont {Malvache}}]{haimerl2019}%
  \BibitemOpen
  \bibfield  {author} {\bibinfo {author} {\bibfnamefont {C.}~\bibnamefont
  {Haimerl}}, \bibinfo {author} {\bibfnamefont {D.}~\bibnamefont
  {Angulo-Garcia}}, \bibinfo {author} {\bibfnamefont {V.}~\bibnamefont
  {Villette}}, \bibinfo {author} {\bibfnamefont {S.}~\bibnamefont
  {Reichinnek}}, \bibinfo {author} {\bibfnamefont {A.}~\bibnamefont {Torcini}},
  \bibinfo {author} {\bibfnamefont {R.}~\bibnamefont {Cossart}}, \ and\
  \bibinfo {author} {\bibfnamefont {A.}~\bibnamefont {Malvache}},\ }\href@noop
  {} {\bibfield  {journal} {\bibinfo  {journal} {Proceedings of the National
  Academy of Sciences}\ }\textbf {\bibinfo {volume} {116}},\ \bibinfo {pages}
  {7477} (\bibinfo {year} {2019})}\BibitemShut {NoStop}%
\bibitem [{\citenamefont {Pietras}\ \emph {et~al.}(2022)\citenamefont
  {Pietras}, \citenamefont {Schmutz},\ and\ \citenamefont
  {Schwalger}}]{pietras2022}%
  \BibitemOpen
  \bibfield  {author} {\bibinfo {author} {\bibfnamefont {B.}~\bibnamefont
  {Pietras}}, \bibinfo {author} {\bibfnamefont {V.}~\bibnamefont {Schmutz}}, \
  and\ \bibinfo {author} {\bibfnamefont {T.}~\bibnamefont {Schwalger}},\
  }\href@noop {} {\bibfield  {journal} {\bibinfo  {journal} {PLOS Computational
  Biology}\ }\textbf {\bibinfo {volume} {18}},\ \bibinfo {pages} {e1010809}
  (\bibinfo {year} {2022})}\BibitemShut {NoStop}%
\bibitem [{\citenamefont {Landau}\ \emph {et~al.}(2016)\citenamefont {Landau},
  \citenamefont {Egger}, \citenamefont {Dercksen}, \citenamefont
  {Oberlaender},\ and\ \citenamefont {Sompolinsky}}]{landau2016}%
  \BibitemOpen
  \bibfield  {author} {\bibinfo {author} {\bibfnamefont {I.~D.}\ \bibnamefont
  {Landau}}, \bibinfo {author} {\bibfnamefont {R.}~\bibnamefont {Egger}},
  \bibinfo {author} {\bibfnamefont {V.~J.}\ \bibnamefont {Dercksen}}, \bibinfo
  {author} {\bibfnamefont {M.}~\bibnamefont {Oberlaender}}, \ and\ \bibinfo
  {author} {\bibfnamefont {H.}~\bibnamefont {Sompolinsky}},\ }\href@noop {}
  {\bibfield  {journal} {\bibinfo  {journal} {Neuron}\ }\textbf {\bibinfo
  {volume} {92}},\ \bibinfo {pages} {1106} (\bibinfo {year}
  {2016})}\BibitemShut {NoStop}%
\bibitem [{\citenamefont {Mongillo}\ \emph {et~al.}(2012)\citenamefont
  {Mongillo}, \citenamefont {Hansel},\ and\ \citenamefont
  {Van~Vreeswijk}}]{mongillo2012}%
  \BibitemOpen
  \bibfield  {author} {\bibinfo {author} {\bibfnamefont {G.}~\bibnamefont
  {Mongillo}}, \bibinfo {author} {\bibfnamefont {D.}~\bibnamefont {Hansel}}, \
  and\ \bibinfo {author} {\bibfnamefont {C.}~\bibnamefont {Van~Vreeswijk}},\
  }\href@noop {} {\bibfield  {journal} {\bibinfo  {journal} {Physical review
  letters}\ }\textbf {\bibinfo {volume} {108}},\ \bibinfo {pages} {158101}
  (\bibinfo {year} {2012})}\BibitemShut {NoStop}%
\bibitem [{\citenamefont {Hansel}\ and\ \citenamefont
  {Mato}(2013)}]{hansel2013}%
  \BibitemOpen
  \bibfield  {author} {\bibinfo {author} {\bibfnamefont {D.}~\bibnamefont
  {Hansel}}\ and\ \bibinfo {author} {\bibfnamefont {G.}~\bibnamefont {Mato}},\
  }\href@noop {} {\bibfield  {journal} {\bibinfo  {journal} {Journal of
  Neuroscience}\ }\textbf {\bibinfo {volume} {33}},\ \bibinfo {pages} {133}
  (\bibinfo {year} {2013})}\BibitemShut {NoStop}%
\end{thebibliography}%

\end{document}